\documentclass[a4paper,11pt]{article}
\pdfoutput=1 

\usepackage{jheppub} 

\usepackage[T1]{fontenc} 

\usepackage[utf8]{inputenc}
\usepackage{amssymb}
\usepackage{amsmath}
\usepackage{amsfonts}
\usepackage{graphicx}
\usepackage{color}
\usepackage{xspace}
\usepackage{comment}
\usepackage{hyperref}
\usepackage[normalem]{ulem}
\usepackage[section]{placeins}
\usepackage{afterpage}
\usepackage{slashed}
\usepackage{enumerate}
\usepackage{enumitem}
\usepackage{caption}
\usepackage{subfigure}
\usepackage{float}
\usepackage{ulem}
\usepackage{verbatim}
\usepackage{multirow,rotating}
\usepackage[dvipsnames]{xcolor}
\usepackage{indentfirst}
\makeatletter
\def\@fpheader{}
\makeatother

\definecolor{dcolour}{rgb}{.5, .5, .5}
\def\gsim{\raise0.3ex\hbox{$\;>$\kern-0.75em\raise-1.1ex\hbox{$\sim\;$}}}
\def\lsim{\raise0.3ex\hbox{$\;<$\kern-0.75em\raise-1.1ex\hbox{$\sim\;$}}}
\def\gsim{\raise0.3ex\hbox{$\;>$\kern-0.75em\raise-1.1ex\hbox{$\sim\;$}}}
\def\lsim{\raise0.3ex\hbox{$\;<$\kern-0.75em\raise-1.1ex\hbox{$\sim\;$}}}

\newcommand{\ba}[1]{\begin{eqnarray} \label{(#1)}}
\newcommand{\ea}{\end{eqnarray}}

\title{Model-independent probes of CP violation in the heavy scalar sector at muon colliders}

\author[a,1]{Qianxi Li \note{Corresponding Author}}
\author[a,1]{, Ying-nan Mao}
\author[a,1]{and Kechen Wang}

\affiliation[a]{Department of Physics, School of Physics and Mechanics, Wuhan University of Technology, Wuhan 430070, Hubei, China}

\emailAdd{qianxi\_li@whut.edu.cn}
\emailAdd{ynmao@whut.edu.cn}
\emailAdd{kechen.wang@whut.edu.cn}

\abstract{
We propose a model-independent test of CP violation in the scalar sector. We consider a heavy neutral scalar $h_2$ with tree-level couplings at the $h_2 V V$ and $h_2 h_1 Z$ vertices (with $V=W^{\pm},Z$), alongside the 125~GeV SM-like Higgs boson $h_1$. At future muon colliders (MuC), we exploit vector-boson-fusion (VBF) production of $h_2$ followed by the decay $h_2 \to Z h_1$. In our framework, observing the single process $V V \to h_2 \to Z h_1$ implies both relevant couplings are nonzero, which is sufficient to establish CP violation in the scalar sector. We simulate signal and backgrounds at $\sqrt{s}=3~(10)$~TeV with integrated luminosity $L=0.9~(10)~\mathrm{ab}^{-1}$.
We then present the expected discovery sensitivities across the $(c_2,c_{12})$ parameter space (with the coupling parameters $c_{2}$ and  $c_{12}$ defined in the text) for multiple $m_{h_2}$ hypotheses.
}

\begin{document}

\maketitle

\flushbottom

\section{Introduction}
\label{sec:intro}

CP violation was first discovered in 1964 in the rare decay of the long-lived kaon, $K_L \to \pi^+\pi^-$~\cite{Christenson:1964fg}. Additional CP-violating effects have since been measured in meson and baryon systems~\cite{Belle:2001zzw,BaBar:2001pki,BaBar:2004gyj,Khalil:2022toi,ParticleDataGroup:2024cfk,LHCb:2025ray}. In the Standard Model (SM), CP violation arises solely from the Kobayashi-Maskawa (KM) mechanism~\cite{Kobayashi:1973fv}, equivalently the complex phase of the Cabibbo-Kobayashi-Maskawa (CKM) matrix~\cite{Cabibbo:1963yz,Kobayashi:1973fv}. The KM mechanism successfully describes all established CP-violating observables~\cite{nir2006cp,HFLAV:2022esi,ParticleDataGroup:2024cfk}. However, it cannot account for the observed matter--antimatter asymmetry of the Universe~\cite{Cohen:1991iu,Cohen:1993nk,Morrissey:2012db}, motivating searches for new sources of CP violation beyond the SM (BSM).

In the SM there is no CP violation originating from the scalar sector. Theoretically, many BSM scenarios introduce additional sources of CP violation, especially in models with an extended scalar sector~\cite{Lee:1974jb,Weinberg:1976hu,Georgi:1978xz,Bento:1991ez,Branco:2011iw}. In such models, the Higgs boson(s) can be CP mixed and thereby exhibit CP-violating interactions. Experimentally, most of these scenarios already face strict constraints from electric dipole moment (EDM) measurements of the electron, neutron, atoms, etc.~\cite{ACME:2018yjb,Roussy:2022cmp,Abel:2020pzs,Graner:2016ses}, because they typically predict EDMs much larger than in the SM~\cite{Pospelov:2005pr,Engel:2013lsa,Yamanaka:2017mef,Safronova:2017xyt,Chupp:2017rkp,Yamaguchi:2020eub,Yamaguchi:2020dsy}. However, EDMs alone cannot distinguish different CP-violation sources, so direct probes of CP-violating interactions at colliders are also essential.\footnote{These two complementary avenues have generated a rich phenomenology of CP violation in BSM scenarios; see, e.g., Refs.~\cite{ElKaffas:2006gdt,Berge:2008wi,Shu:2013uua,Mao:2014oya,Chen:2015gaa,Keus:2015hva,Li:2015kxc,Shen:2015pha,Fontes:2015mea,Bian:2016awe,Mao:2016jor,Chen:2017com,Mao:2017hpp,Chen:2017plj,Wan:2017qiq,Mao:2018kfn,Cheung:2020ugr,Cao:2020hhb,Azevedo:2020fdl,Azevedo:2020vfw,Pan:2020qqd,Antusch:2020ngh,Kanemura:2021atq,Low:2020iua,Enomoto:2021dkl,Enomoto:2022rrl,Liu:2023gwu,Cheung:2023qnj,Hou:2023kho,Boto:2024jgj,Biekotter:2024ykp,Enomoto:2024jyc,Aiko:2025tbk,Davila:2025goc,Boto:2025ovp,Lee:2025nuj} for details.}

The discovery of the 125~GeV SM-like Higgs boson (denoted $h_1$ below) at the Large Hadron Collider (LHC)~\cite{Aad:2012tfa,CMS:2012qbp,Aad:2015zhl} marked a turning point in particle physics, making tests of its CP properties crucial for elucidating the Higgs sector. Current data are consistent with a purely CP-even $h_1$, but a CP-mixed $h_1$ is not excluded~\cite{Cheung:2020ugr}. In models with an extended scalar sector, there can be additional states $h_{2,3,4,\ldots}$ ($h_i$ with $i>1$), whose possible CP mixing leads to rich collider signatures.

As a representative example, CP violation in the Yukawa sector has been widely studied. For a CP-mixed scalar $h_i$ coupled to fermions via
\begin{equation}
\mathcal{L}\supset -\, h_i\,\bar{f}\,(g_S + i g_P \gamma^{5})\,f,
\end{equation}
one typically considers $f=t$~\cite{Schmidt:1992et,Mahlon:1995zn,Asakawa:2003dh,BhupalDev:2007ftb,He:2014xla,Boudjema:2015nda,Buckley:2015vsa,Hagiwara:2016rdv,AmorDosSantos:2017ayi,Azevedo:2017qiz,Bernreuther:2017cyi,Hagiwara:2017ban,Ma:2018ott,Cepeda:2019klc,Faroughy:2019ird,Cheung:2020ugr,Cao:2020hhb,Azevedo:2020vfw,Azevedo:2020fdl,Cheung:2023qnj,Barger:2023wbg,Cassidy:2023lwd,Esmail:2024gdc,Hammad:2025ewr} or $f=\tau$~\cite{Desch:2003rw,Berge:2008wi,Harnik:2013aja,Berge:2013jra,Berge:2014sra,Berge:2015nua,Askew:2015mda,Jozefowicz:2016kvz,Hagiwara:2016zqz,Antusch:2020ngh,Kanemura:2021atq,Kanemura:2021dez}, since the spin information of $t$ or $\tau$ is imprinted in final-state distributions and is sensitive to the CP nature of $h_i$. Observation of both CP-even and CP-odd components of $h_i$ would confirm CP violation. Such an idea to test CP violation is effective for both the 125 GeV Higgs boson ($i=1$) or an additional scalar ($i>1$).

CP violation can also be probed purely in the bosonic sector through interactions of scalars with the massive SM gauge bosons $W^\pm$ and $Z$, as proposed in Ref.~\cite{Li:2016zzh}. To employ this method at tree level, at least two scalars must be discovered; the effective interaction can be written as~\cite{Li:2016zzh}
\begin{equation}
\label{eq:effintgeneral}
\mathcal{L}\supset\left(\frac{g^2 v}{2}W^{+\mu}W_\mu^-+\frac{g^2 v}{4c_W^2}Z^{\mu}Z_\mu\right)\left(\sum_i c_i h_i\right)
+\sum_{i<j}\frac{c_{ij}g}{2c_W}Z_\mu\left(h_i\partial^\mu h_j-h_j\partial^\mu h_i\right),
\end{equation}
where $g$ is the weak gauge coupling, $v=246~\mathrm{GeV}$ is the vacuum expectation value (VEV) of the SM Higgs doublet, and $c_W\equiv\cos\theta_W$ is the cosine of the Weinberg angle $\theta_W$. The parameters $c_i$ denote the $h_i V V$ coupling strengths relative to the SM; and for the SM-like Higgs boson $h_1$, we have $c_1\simeq 1$. For any two distinct scalars $h_i$ and $h_j$, the quantity $K\equiv c_i c_j c_{ij}$ serves as a measure of CP violation \cite{Mendez:1991gp,Branco:2011iw,Li:2016zzh}. Since nonzero $c_i$ implies a CP-even component in $h_i$, and nonzero $c_{ij}$ implies that $h_i$ and $h_j$ contain components of opposite CP number, $K\neq 0$ is a sufficient condition to establish CP violation. Recent collider studies employing this method include Refs.~\cite{Li:2016zzh,Mao:2017hpp,Mao:2018kfn,Capucha:2025qgr}.

In this paper we focus on CP violation in the pure bosonic sector following Ref.~\cite{Li:2016zzh}. Ref.~\cite{Li:2016zzh} considers a light extra scalar $h_2$ and prospects at a Higgs factory with $\sqrt{s}=250$~GeV (e.g., CEPC~\cite{CEPCStudyGroup:2018ghi,CEPCStudyGroup:2023quu,Ai:2025cpj}), however, light new states are already tightly constrained; we therefore study a heavy $h_2$ scenario with $m_{h_2}\sim\mathcal{O}(\mathrm{TeV})$. We consider future muon colliders (MuC)~\cite{Delahaye:2019omf,Aime:2022flm,Black:2022cth,Accettura:2023ked,InternationalMuonCollider:2024jyv,InternationalMuonCollider:2025sys,USMCC:2025WP} with $\sqrt{s}$ up to $10~\mathrm{TeV}$, where TeV-scale scalars are kinematically accessible. Specifically, we study vector-boson-fusion (VBF) production with $h_2\to Z h_1$, i.e., $V V\to h_2\to Z h_1$ with $V=W^\pm, Z$. Observation of this single process implies $c_2\neq 0$ and $c_{12}\neq 0$ simultaneously, thereby probing CP violation in the scalar sector.

The paper is organized as follows. In Sec.~\ref{sec:method} we describe our model-independent method to search for CP violation. In Sec.~\ref{sec:simulation} we present the simulation setup and analysis. In Sec.~\ref{sec:results} we show the results. Conclusions are given in Sec.~\ref{sec:conclusion}, with additional details in the Appendices \ref{app:sumrule} -- \ref{app:cuts}.

\section{Model-independent method}
\label{sec:method}

\begin{figure}[h]
\centering
\includegraphics[width=0.4\linewidth]{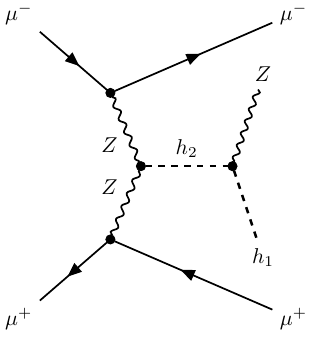}
\hspace{+6mm}
\includegraphics[width=0.4\linewidth]{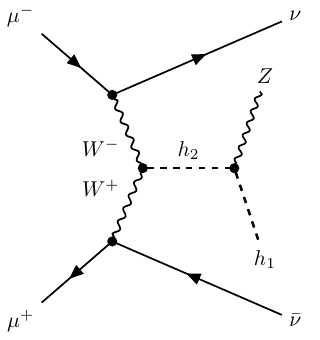}
\caption{Feynman diagrams for the VBF processes $V V \to h_2 \to Z h_1$ (with $V = W^{\pm}, Z$).}
\label{fig:Feynman diagrams}
\end{figure}

Assuming that a heavy neutral scalar $h_2$ has been discovered, following Ref.~\cite{Li:2016zzh}, we write the tree level effective interaction as
\begin{equation}
\label{eq:effint}
\mathcal{L} \supset \left(\frac{g^2 v}{2} W^{+\mu} W^-_\mu + \frac{g^2 v}{4 c_W^2} Z^{\mu} Z_\mu\right)\left(c_1 h_1 + c_2 h_2\right)
+ \frac{c_{12} g}{2 c_W} Z_\mu \left(h_1 \partial^\mu h_2 - h_2 \partial^\mu h_1\right),
\end{equation}
without further assumptions on the ultraviolet (UV) complete model. Since $c_1 \simeq 1$, CP violation can be confirmed if both $c_2$ and $c_{12}$ are nonzero, as discussed above. It is then natural to study the VBF processes $V V \to h_2 \to Z h_1$ with $V = W^{\pm}, Z$ at MuC, whose Feynman diagrams are shown in Fig.~\ref{fig:Feynman diagrams}.

\begin{figure}[h]
\centering
\includegraphics[width=0.45\linewidth]{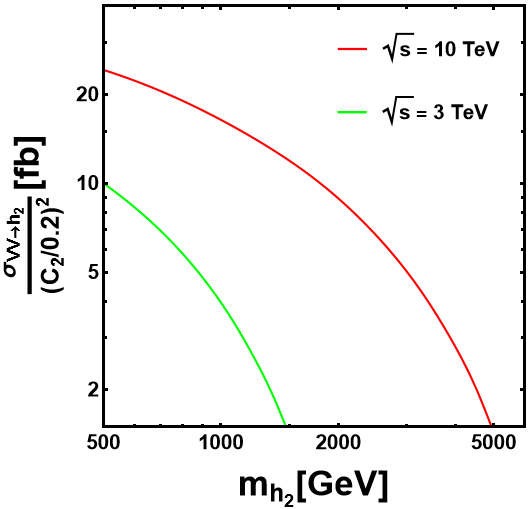}
\hspace{+6mm}
\includegraphics[width=0.45\linewidth]{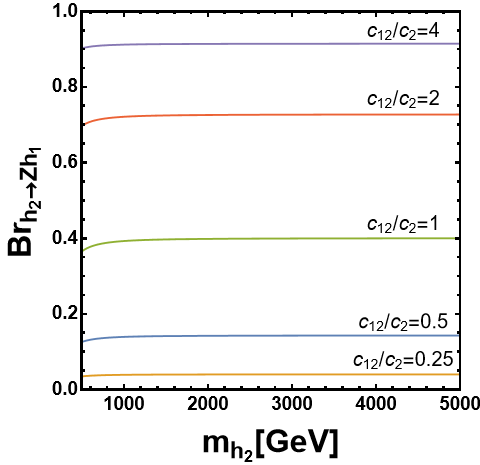}
\caption{Left: total cross sections $\sigma$ of $V V \to h_2$ at future MuC with $\sqrt{s} = 3$ and $10$~TeV as a function of $m_{h_2}$. Right: $\mathrm{Br}_{h_2\to Z h_1}$ as a function of $m_{h_2}$ for different $c_{12}/c_2$.}
\label{fig:section and br}
\end{figure}

In these processes, $h_2$ is produced via the $h_2 V V$ vertex, giving a total cross section $\sigma_{VV\rightarrow h_2}\propto c_2^2$, and subsequently decays through the $h_1 h_2 Z$ vertex with branching ratio $\mathrm{Br}_{h_2\to Z h_1} \equiv \Gamma_{h_2\to Z h_1}/\Gamma_{h_2} \propto c_{12}^2/\Gamma_{h_2}$, where $\Gamma_{h_2}$ denotes the total decay width of $h_2$.
If either $c_2 = 0$ or $c_{12} = 0$, then $\sigma_{VV\rightarrow h_2} \cdot \mathrm{Br}_{h_2\to Z h_1} = 0$ and the process is absent; conversely, the observation of such a process implies $c_2 c_{12} \neq 0$ and hence CP violation in the scalar sector.
We compute $\sigma_{VV\rightarrow h_2}$ at future MuC with $\sqrt{s} = 3$ and $10$~TeV, respectively, using \texttt{MadGraph5\_aMC@NLO} (v3.6.2)~\cite{Alwall:2014hca}, and show its dependence on $m_{h_2}$ in Fig.~\ref{fig:section and br} (left). For given $m_{h_2}$ and $\sqrt{s}$, the cross section of $W^+W^-$-fusion is about an order larger than that of $ZZ$-fusion, which means that $W^+W^-$-fusion is the dominant process. For $c_2 \simeq 0.2$ which is a typical benchmark point in the following analysis, the total cross section $\sigma_{VV\rightarrow h_2} \propto c_2^2$ can reach at most $\mathcal{O}(1\text{--}10)\,\mathrm{fb}$ in the mass region considered in this work, and decreases quickly as $m_{h_2}$ increases.

We assume that Eq.~(\ref{eq:effint}) contains all the interactions related to $h_2$, and thus $h_2$ decays only through the following three channels: $h_2 \to Z h_1$, $h_2 \to Z Z$, and $h_2 \to W^+ W^-$. The branching ratio for the decay channel $h_2 \to Z h_1$ is defined as
\begin{equation}
\label{eq:BRh2zh1}
\mathrm{Br}_{h_2\to Z h_1} \equiv \frac{\Gamma_{h_2\to Z h_1}}{\Gamma_{h_2}}
= \frac{\Gamma_{h_2\to Z h_1}}{\Gamma_{h_2\to Z h_1} + \Gamma_{h_2\to W^+W^-} + \Gamma_{h_2\to Z Z}},
\end{equation}
where the partial decay widths are given as~\cite{Branco:2011iw,Djouadi:2005gi,Djouadi:2005gj}
\begin{eqnarray}
\label{eq:widthh2ww}
\Gamma_{h_{2}\to W^{+}W^{-}} &=& \frac{m_{h_{2}}^{3} c_2^{2}}{16\pi v^{2}} \sqrt{1-\frac{4m_{W}^{2}}{m_{h_{2}}^{2}}}\left(1-\frac{4m_{W}^{2}}{m_{h_{2}}^{2}}+\frac{12m_{W}^{4}}{m_{h_{2}}^{4}}\right),\\
\label{eq:widthh2zz}
\Gamma_{h_2\to Z Z} &=& \frac{m_{h_2}^{3} c_2^2}{32\pi v^2} \sqrt{1-\frac{4m_Z^2}{m_{h_2}^2}}\left(1-\frac{4m_Z^2}{m_{h_2}^2}+\frac{12m_Z^4}{m_{h_2}^4}\right),\\
\label{eq:widthh2zh1}
\Gamma_{h_2\to Z h_1} &=& \frac{m_{h_2}^3 c_{12}^2}{16\pi v^2}\, f^3\!\left(\frac{m_Z^2}{m_{h_2}^2},\frac{m_{h_1}^2}{m_{h_2}^2}\right).
\end{eqnarray}
Here $m_{h_1,h_2,Z,W}$ denote the corresponding particle masses, and the function
\begin{equation}
\label{eq:function}
f(x,y) \equiv \sqrt{1 + x^2 + y^2 - 2x - 2y - 2xy}.
\end{equation}
If $m_{h_2} \gg m_{h_1,W,Z}$, the total decay width of $h_2$ approximately behaves as
\begin{equation}
\label{eq:totalwidth}
\Gamma_{h_2} \simeq \left(493\,c_2^2 + 329\,c_{12}^2\right) \left(\frac{m_{h_2}}{1\,\mathrm{TeV}}\right)^3 \,\mathrm{GeV}.
\end{equation}
Numerically, Fig.~\ref{fig:section and br} (right) shows $\mathrm{Br}_{h_2\to Z h_1}$ as a function of $m_{h_2}$ for different ratios $c_{12}/c_2$.
The branching ratio is sensitive to $c_{12}/c_2$ but only weakly dependent on $m_{h_2}$.
In particular, in the limit $m_{h_2} \gg m_{h_1,W,Z}$, we have approximately
\begin{equation}
\mathrm{Br}_{h_2\to Z h_1} \simeq \frac{2(c_{12}/c_2)^2}{2(c_{12}/c_2)^2 + 3},
\end{equation}
which is nearly independent on $m_{h_2}$.

\section{Simulation studies}
\label{sec:simulation}

\subsection{Simulation setup}
In this work, we study the VBF processes
$V V \to h_2 \to Z(\to \ell^+\ell^-)\, h_1(\to b\bar b)$ with $V=W^{\pm}, Z$ and $\ell = e, \mu$ at MuC, assuming that a heavy $h_2$ has already been discovered. The effective interaction in Eq.~\eqref{eq:effint} is implemented with \texttt{FeynRules} (v2.0) \cite{Alloul:2013bka} to produce a Universal FeynRuls Output (UFO) model, which we load in \texttt{MadGraph5\_aMC@NLO} (v3.6.2)~\cite{Alwall:2014hca}
to simulate the $\mu^-\mu^+$ collisions, and generate events for the whole processes of both signal and backgrounds.
Event generation is then interfaced to \texttt{PYTHIA}~8.3~\cite{Bierlich:2022pfr}
for parton showering and hadronization. Detector response is simulated with \texttt{Delphes}~3.4.2~\cite{deFavereau:2013fsa}
using a muon-collider configuration card.

An absorber is installed at MuC to mitigate beam-induced backgrounds (BIB),
so only objects within $10^{\circ}<\theta<170^{\circ}$ are reconstructed, where
$\theta$ denotes the polar angle with respect to the $z$-axis (the initial $\mu^-$ beam direction)~\cite{Mokhov:2011zzd,Mokhov:2014hza,DiBenedetto:2018cpy,Bartosik:2019dzq}.
Objects produced in the very forward/backward regions are absorbed and thus considered to be undetected in this study.

We must treat jets in this work, because the signal we consider here includes two $b$-jets, and we must also consider
the background processes including $b$-jets or light jets which are mis-tagged as $b$-jets.
For jet reconstruction, we use the algorithm of ``VLCjetR05N2''~\cite{Cacciari:2011ma,Boronat:2014hva,Boronat:2016tgd,CLICdp:2018vnx}.
The double-$b$-tagging efficiency is approximately $0.8$, while the probability to mis-tag non-$b$-jets
as two $b$-jets can reach about $\mathcal{O}(10^{-2})$~\cite{CLICdp:2018vnx}.

\subsection{Signal and background processes}
At future MuC, we target the VBF processes $V V \to h_2 \to h_1 Z$ to search for CP violation in the scalar sector,
as shown in Fig.~\ref{fig:Feynman diagrams}.
For most events from the $Z Z$-fusion topology (left panel of Fig.~\ref{fig:Feynman diagrams}),
the scattered muons emerge at very small polar angles and low transverse momentum $p_T$,
and therefore are outside the detector acceptance. We select $Z \to \ell^-\ell^+$ (with $\ell=e,\mu$) and $h_1 \to b\bar b$.
For $W^+W^-$-fusion, the final state contains two energetic neutrinos, which are undetectable; while for $ZZ$-fusion, the final state contains two energetic $\mu^\pm$, which appear in the very forward/backward region in most events, and thus cannot be detected as well. Hence we generate the two VBF topologies together, in which the final state contains large missing energy, two opposite-charge, same-flavor leptons, and two $b$-tagged jets.

The signal-generation cross section $\sigma_{\mathrm{sig}}$ scales with a parameter
\begin{equation}
\label{eq:kappa}
\kappa \equiv c_2^2 \cdot \mathrm{Br}_{h_2\to Z h_1},
\end{equation}
and thus depends on $m_{h_2}$, $c_2$, and $c_{12}$. In the limit $m_{h_2} \gg m_{h_1,W,Z}$, we have approximately $\kappa\simeq\frac{2c_2^2c_{12}^2}{3c_2^2+2c_{12}^2}$ based on the results in Eqs. (\ref{eq:widthh2ww}) -- (\ref{eq:widthh2zh1}).
For the chosen final state, the cross section is also proportional to the SM branching fractions~\cite{ParticleDataGroup:2024cfk}
\begin{equation}
\label{eq:BRzll and eq:BRh1bb}
\mathrm{Br}_{Z \to \ell^-\ell^+}=0.033647,
\quad\quad\mathrm{and}\quad\quad
\mathrm{Br}_{h_1 \to b\bar b}=0.5824.
\end{equation}
Global fits to the 125~GeV Higgs signal strengths data show that $c_1\gtrsim0.95$ at $95\%$ C.L.~\cite{Cheung:2020ugr} assuming no exotic $h_1$ decay channels~\footnote{The global fits in \cite{Cheung:2020ugr} were performed based on two-Higgs-doublet model (2HDM), but the data strongly favor the SM-like limit, and hence the constraint on $c_1$ depends weakly on the specific UV complete model assumptions. On the other hand, $c_1\simeq0.95$ is still allowed by LHC data.}. Even under the assumption of custodial symmetry in Higgs sector \cite{ParticleDataGroup:2024cfk}, the sum rule on the couplings $c_1,~c_2$, and $c_{12}$ still depends on the UV complete model construction, especially the weak hypercharge of the extended scalar sector. In Appendix \ref{app:sumrule}, we derive it in a simplified scenario. Following the sum rule, we typically choose the parameter region as
$c_2^2+c_{12}^2 \lesssim 1-c_1^2 \lesssim 0.1$, which is still allowed by current LHC data; and thus for the phenomenological studies, we choose the parameter region $c_2 \le 0.2$ and $c_{12} \le 0.2$ as benchmarks \footnote{In this paper, we only discuss the case $c_2,c_{12}>0$ for simplicity.}. Direct LHC searches for a heavy scalar decaying to $Z Z$ final state further constrain $c_2$ for $m_{h_2}\!\lesssim\!700$~GeV~\cite{ATLAS:2020tlo}; and thus we choose all the benchmark points with $m_{h_2} \gtrsim 700$~GeV due to the consistency with LHC bounds~\footnote{Note that the direct LEP bounds on $(c_2,c_{12})$ quoted in Ref.~\cite{Li:2016zzh} were derived for a light scalar (e.g., $m_2=40~\mathrm{GeV}$), which is kinematically accessible at LEP (and assuming scalar decays dominantly to $b\bar b$), and hence do not apply to our TeV-scale $h_2$ benchmarks.}.

We consider two center-of-mass energies of the $\mu^-\mu^+$ collisions, $\sqrt{s}=3$ and $10$~TeV. The corresponding integrated luminosities are
$L=0.9$ and $10~\mathrm{ab}^{-1}$, respectively~\cite{Delahaye:2019omf}. To generate the signal samples, we choose the following benchmark points:
\begin{equation}
m_{h_2}\in\left\{\begin{array}{ll}\{700,1000\}~\mathrm{GeV},&\quad(\mathrm{for}~\sqrt{s}=3~\mathrm{TeV});\\
\{700,1000,1500,2000,3000,4500\}~\mathrm{GeV},&\quad(\mathrm{for}~\sqrt{s}=10~\mathrm{TeV}).\end{array}\right.
\end{equation}
All signal events are generated with the parameter choice $c_2=c_{12}=0.2$, which leads to the largest allowed width of $h_2$, to ensure that our search strategy works in the whole parameter region with $(c_2,c_{12})\le0.2$.

Given the final state containing undetected forward/backward muons or neutrinos, two opposite-charge same flavor leptons, and two $b$-tagged jets (including the mis-tagged non-$b$-jets), we consider four main backgrounds:
\begin{itemize}
    \item[(i)] $V V \to h_2 \to Z(\to \ell^-\ell^+) Z(\to jj)$ which is a typical BSM background process, labeled “b1” with cross section $\sigma_{\mathrm{b1}}$. Here $\sigma_{\mathrm{b1}}$ scales with a parameter
    \begin{equation}
    \label{eq:xi}
    \xi \equiv c_2^2 \cdot \mathrm{Br}_{h_2\to Z Z},
    \end{equation}
    where
    \begin{equation}
    \label{eq:BRh2zz}
    \mathrm{Br}_{h_2\to Z Z}\equiv\frac{\Gamma_{h_2\to Z Z}}{\Gamma_{h_2}}=\frac{\Gamma_{h_2\to Z Z}}{\Gamma_{h_2\to Z h_1}+\Gamma_{h_2\to W^+W^-}+\Gamma_{h_2\to Z Z}}.
    \end{equation}
    In the limit $m_{h_2} \gg m_{h_1,W,Z}$, we have approximately $\xi\simeq\frac{c_2^4}{3c_2^2+2c_{12}^2}$ based on the results in Eqs. (\ref{eq:widthh2ww}) -- (\ref{eq:widthh2zh1}).
    \item[(ii)] $\mu^-\mu^+ \to \nu \bar{\nu}\, Z(\to \ell^-\ell^+)\, h_1(\to b\bar b)$ and $\mu^-\mu^+ \to \ell^-\ell^+\, Z(\to \ell^-\ell^+)\, h_1(\to b\bar b)$, labeled “b2” with cross section $\sigma_{\mathrm{b2}}$.
    \item[(iii)] $\mu^-\mu^+ \to \nu \bar{\nu}\, Z(\to \ell^-\ell^+)\, Z(\to jj)$ and $\mu^-\mu^+ \to \ell^-\ell^+\, Z(\to \ell^-\ell^+)\, Z(\to jj)$, labeled “b3” with cross section $\sigma_{\mathrm{b3}}$.
    \item[(iv)] $\mu^-\mu^+ \to b \bar b \, W^-(\to \bar\nu \ell^-)\, W^+(\to \nu \ell^+)$, labeled “b4” with cross section $\sigma_{\mathrm{b4}}$.
\end{itemize}
The total cross section of backgrounds is defined as $\sigma_{\mathrm{bkg}}\equiv\sigma_{\mathrm{b1}}+\sigma_{\mathrm{b2}}+\sigma_{\mathrm{b3}}+\sigma_{\mathrm{b4}}$.

Background events are generated with the same tools and settings as the signal.
To maintain consistency, event yields are estimated from the \texttt{MadGraph5\_aMC@NLO} (v3.6.2)~\cite{Alwall:2014hca}
production cross sections. Note that b1 is a BSM background process, whose cross section depends on the parameters $m_{h_2},~c_2$, and $c_{12}$; while b2-b4 are all SM background processes, whose cross sections do not depend on the BSM parameters.

\subsection{Analysis strategy}
\subsubsection{Pre-selection}
Final-state leptons and $b$-jets are ordered by descending transverse momentum and labeled $O_i$ ($i=1,2,\ldots$), with $O\in\{\ell,\,b\text{-jet}\}$. We apply the following pre-selection cuts to select the target final state and reject background at the first stage:
\begin{itemize}
    \item[(i)] Exactly two opposite-charge, same-flavor leptons, i.e. $N(\ell)=2$ with $\ell=e,\mu$.
    \item[(ii)] Exactly two $b$-tagged jets, i.e. $N(j_b)=2$.
\end{itemize}
Jets do not carry a well-defined electric charge; the so-called ``jet charge'' is an estimator and is not used here to require ``opposite charges'' for $b$-jets~\cite{ATLAS:2015rlw,CMS:2017yer,Kogler:2018hem}. The second condition will miss the events in which two $b$-jets are not fully tagged, and include the events in which light jets are mis-tagged as two $b$-jets, as discussed above.

\subsubsection{Reconstruction}
In the signal topology, the heavy $h_2$ decays to one SM Higgs $h_1$ and one $Z$ boson; subsequently $Z\to\ell^+\ell^-$ and $h_1\to b\bar b$. We reconstruct the $h_2$ mass from the invariant mass of the $\ell^+\ell^-b\bar b$ system, $m_{b\bar b\ell^-\ell^+}$; and the $h_1$ mass from the invariant mass of the two $b$ jets, $m_{b\bar b}$. The two undetected energetic forward/backward objects (neutrinos or absorbed muons) form the invisible system, and we define its invariant mass as $m\mkern-10.5mu/$, which captures the VBF kinematic features in both signal and some background processes. In our detailed study, we find that setting selection cuts on $m_{b\bar b}$ and $m_{b\bar b\ell^-\ell^+}$ is enough to separate signal and background events, and thus $m\mkern-10.5mu/$ is only used as a tagging condition for VBF processes.

In this section, we choose two benchmark points: (i)~$m_{h_2}=700~\mathrm{GeV}$ with $\sqrt{s}=3~\textrm{TeV}$, and (ii)~$m_{h_2}=3000~\mathrm{GeV}$ with $\sqrt{s}=10~\textrm{TeV}$, to show how the selection cuts are set for the observables, $m_{b\bar b}$ and $m_{b\bar b\ell^-\ell^+}$, together with the total cross sections for both signal and background after the selection cuts. Details for the other six benchmark points are put in the appendices.

In Fig.~\ref{fig:distrBench3TeV}, we choose the benchmark point (i), and show the normalized distributions of $m_{b\bar b}$, $m_{b\bar b\ell^-\ell^+}$, and $m\mkern-10.5mu/$, for both signal and background processes after pre-selection.
\begin{figure}[h]
\centering
\includegraphics[width=0.33\linewidth]{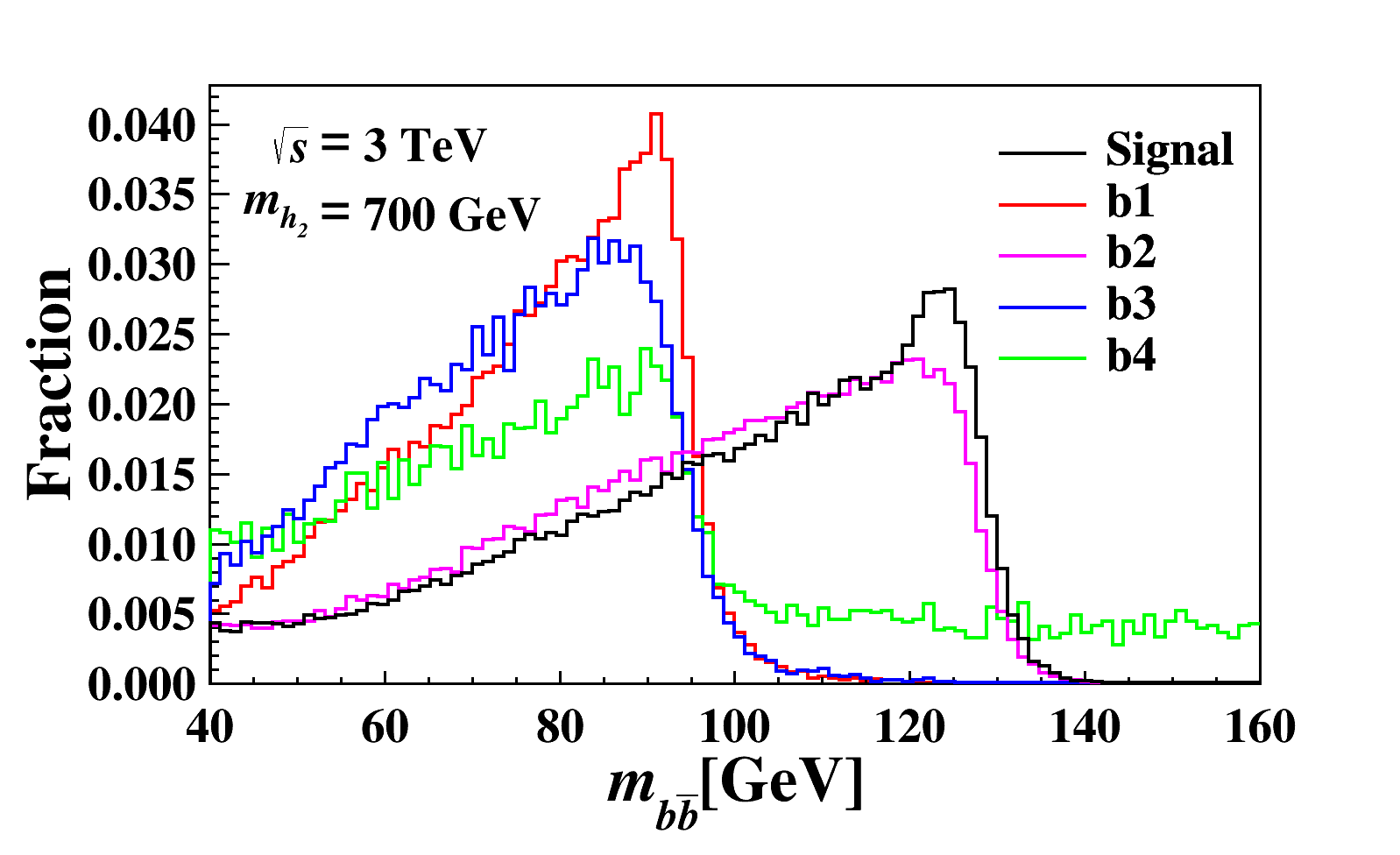}
\hspace{-6mm}
\includegraphics[width=0.33\linewidth]{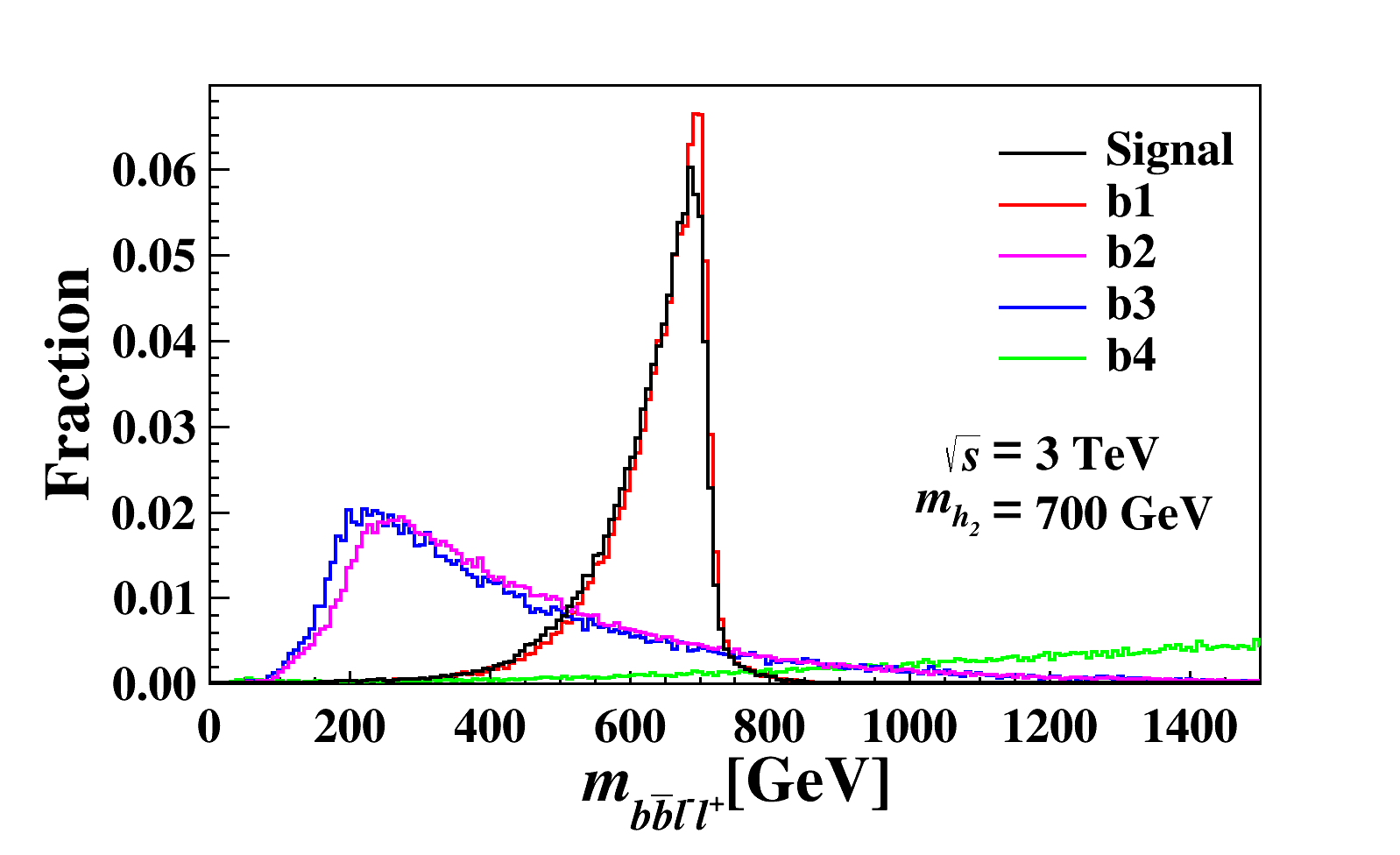}
\hspace{-6mm}
\includegraphics[width=0.33\linewidth]{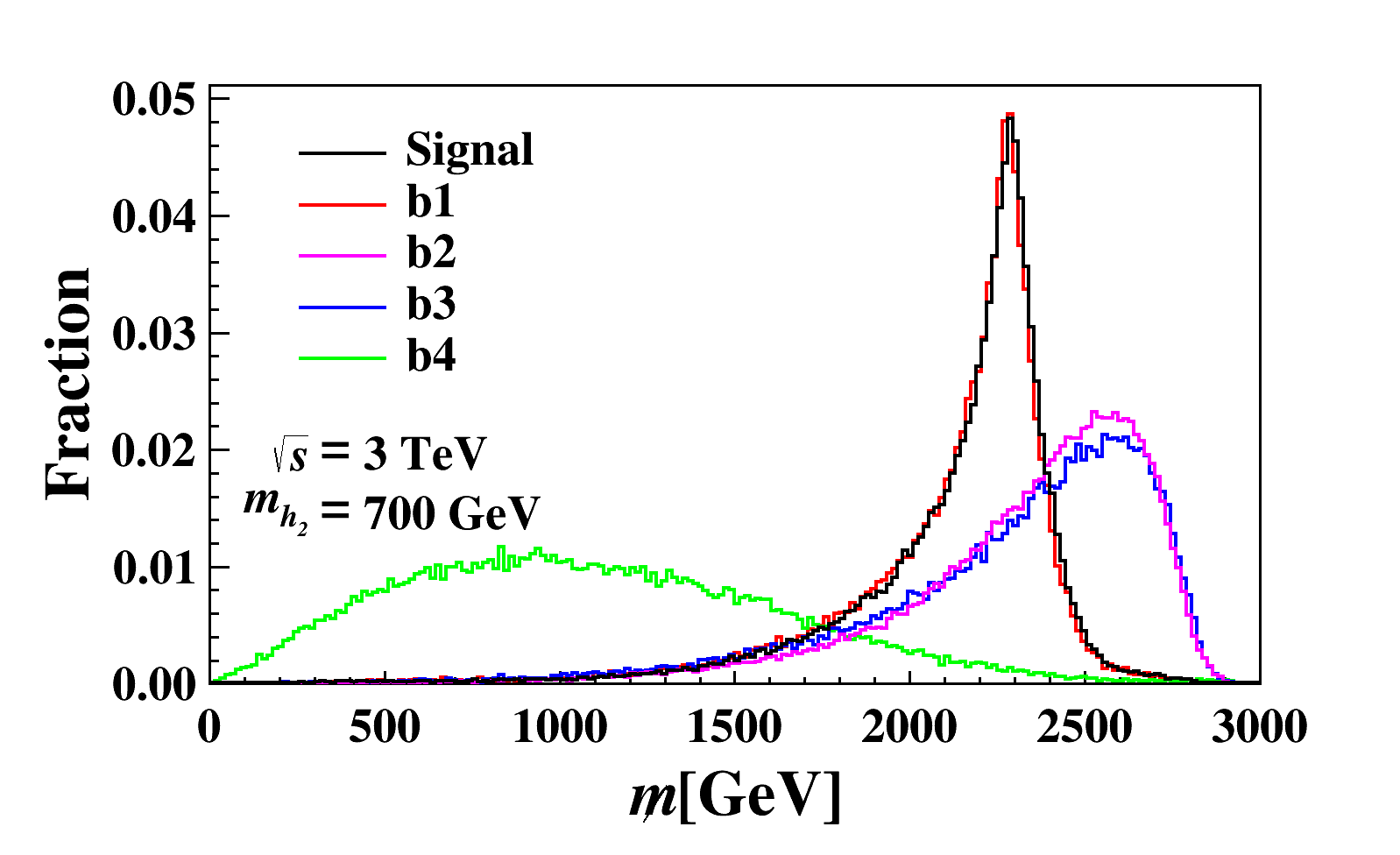}
\caption{
Normalized distributions of $m_{b\bar b}$ (left), $m_{b\bar b\ell^-\ell^+}$ (middle), and $m\mkern-10.5mu/$ (right), for both signal and background processes after pre-selection ($m_{h_2}=700~\mathrm{GeV}$ with $\sqrt{s}=3~\mathrm{TeV}$).
}
\label{fig:distrBench3TeV}
\end{figure}
Based on the left plot, we choose the selection cut $100~\textrm{GeV}\le m_{b\bar{b}}\le 140~\textrm{GeV}$, which is effective to suppress backgrounds b1, b3, and b4. In background b2, the $b\bar{b}$-pairs also come from $h_1$, meaning that its distribution should be similar to that of the signal process. Based on the middle plot, we choose the selection cut $640~\textrm{GeV}\le m_{b\bar b\ell^-\ell^+}\le 720~\textrm{GeV}$ \footnote{The minimal necessary width of the cut window, which relate closely to the dispersion in $m_{b\bar b\ell^-\ell^+}$ distribution, is sensitive to both the energy smearing at detector level and the total width of $h_2$ (especially for large $m_{h_2}$), denoted as $\Gamma_{h_2}$, which is calculated in Eq.~(\ref{eq:totalwidth}). In this work, all signal events are generated using the benchmark point $c_2=c_{12}=0.2$ leading to the largest allowed $\Gamma_{h_2}$, which ensures our search strategy works in the whole parameter region as we consider.}, which is effective to suppress backgrounds b2, b3, and b4. Since b1 is a BSM background which also comes from the $h_2$ decay, the $m_{b\bar b\ell^-\ell^+}$ distribution in b1 should be similar to that of the signal process. If we perform these two selection cuts together, it is efficient to suppress all the four background processes. We also show the distribution of $m\mkern-10.5mu/$ distributions as a VBF tagging condition; however, it cannot provide further separation between the signal and background processes, after we set selection cuts on $m_{b\bar{b}}$ and $m_{b\bar b\ell^-\ell^+}$. After selection cuts, we have the total cross sections for both signal and background processes as
\begin{equation}
\sigma_{\mathrm{sig}}=1.58\kappa~(\mathrm{fb}),\quad\quad\mathrm{and}\quad\quad\sigma_{\mathrm{bkg}}=(14\xi+5.1)\times10^{-3}~(\mathrm{fb});
\end{equation}
which can be used to predict the discovery potential of this process for given model parameters $(c_2,c_{12})$ and integrated luminosity $L$, with $m_{h_2}=700~\textrm{GeV}$.

\begin{figure}[h]
\centering
\includegraphics[width=0.33\linewidth]{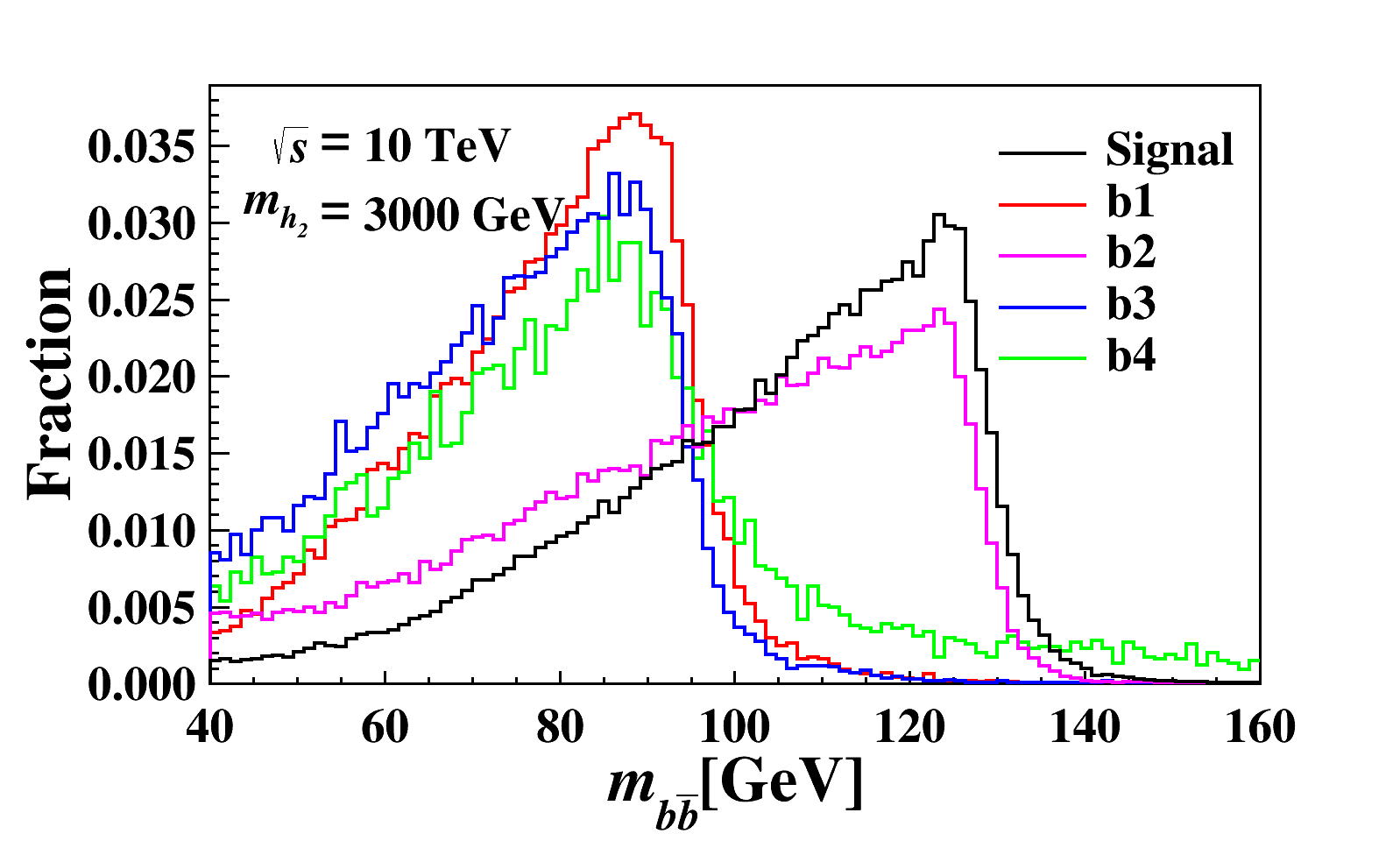}
\hspace{-6mm}
\includegraphics[width=0.33\linewidth]{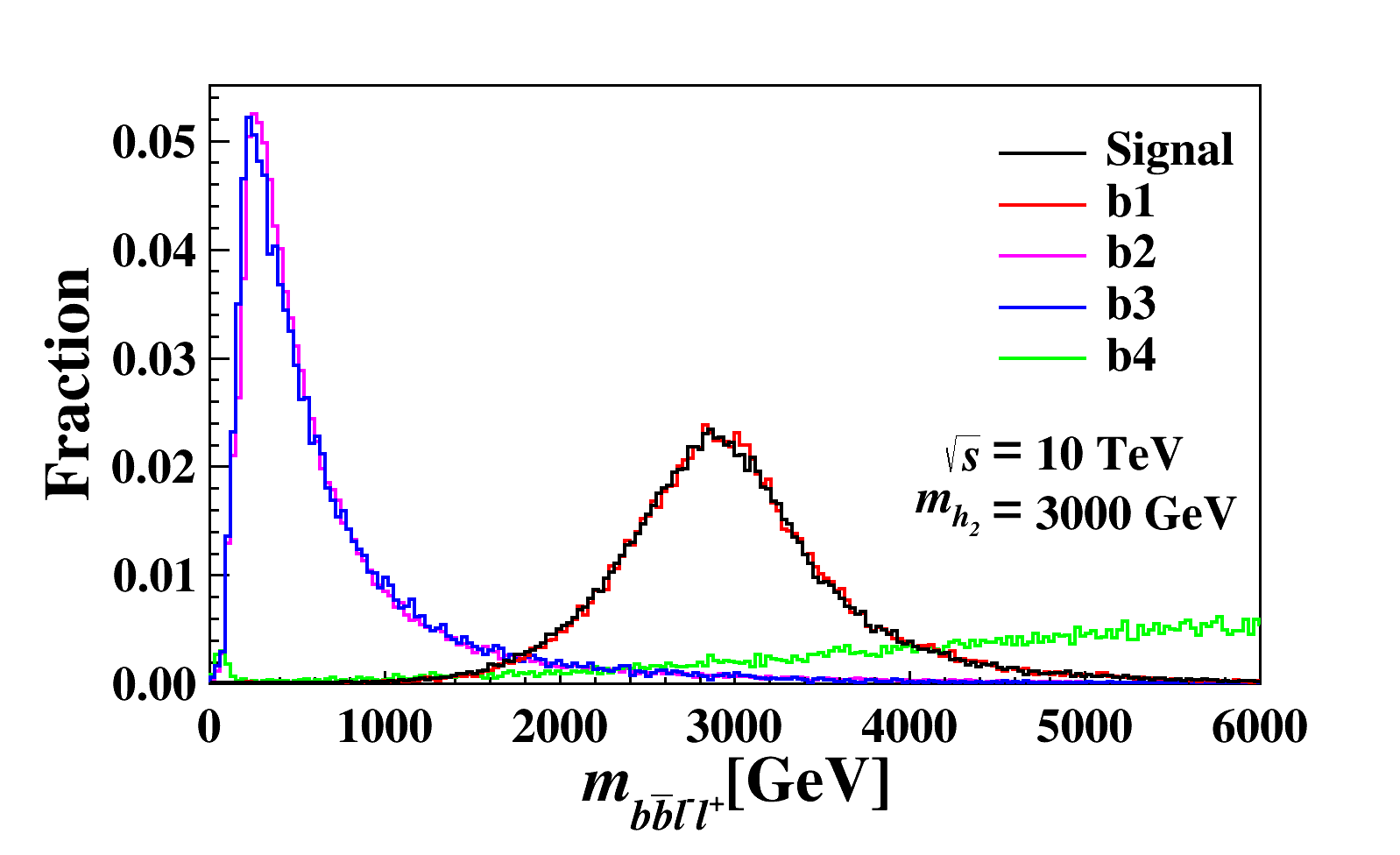}
\hspace{-6mm}
\includegraphics[width=0.33\linewidth]{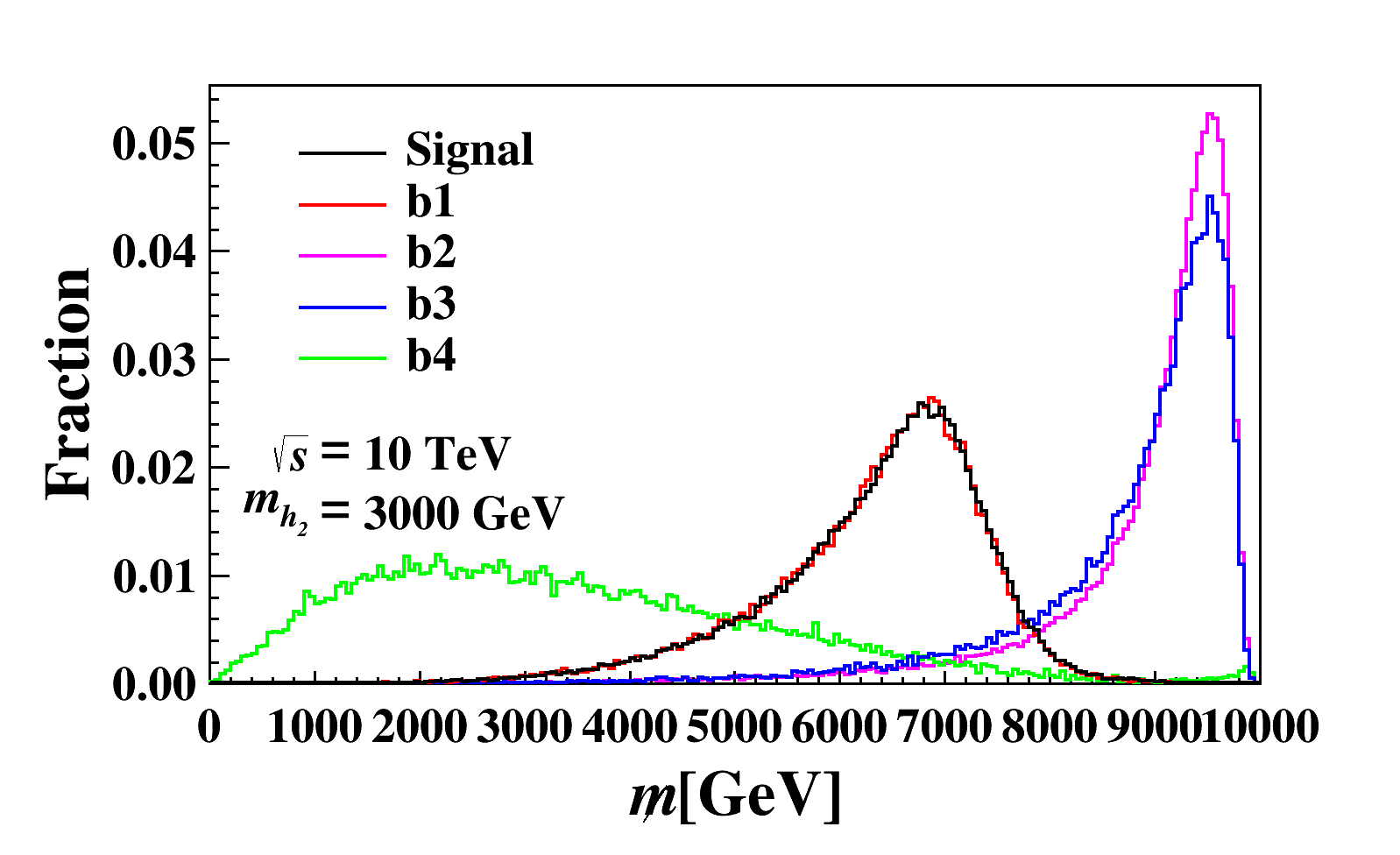}
\caption{
Normalized distributions of $m_{b\bar b}$ (left), $m_{b\bar b\ell^-\ell^+}$ (middle), and $m\mkern-10.5mu/$ (right), for both signal and background processes after preselection ($m_{h_2}=3000~\mathrm{GeV}$ with $\sqrt{s}=10~\mathrm{TeV}$).
}
\label{fig:distrBench10TeV}
\end{figure}

In Fig.~\ref{fig:distrBench10TeV}, we choose the benchmark point (ii), and show the normalized distributions of $m_{b\bar b}$, $m_{b\bar b\ell^-\ell^+}$, and $m\mkern-10.5mu/$, for both signal and background processes after pre-selection.
Based on the left plot, the distributions of $m_{b\bar{b}}$ for both signal and background processes are quite similar to those in the case of benchmark point (i), so thus we choose the selection cut $100~\textrm{GeV}\le m_{b\bar{b}}\le 140~\textrm{GeV}$ again. Based on the middle plot, it can be found that the signal peak is much wider than that for benchmark point (i), and we choose the selection cut $2400~\textrm{GeV}\le m_{b\bar b\ell^-\ell^+}\le 3400~\textrm{GeV}$. Cuts on $m\mkern-10.5mu/$ in the right plot cannot provide further separation between signal and background processes after selection cuts on $m_{b\bar{b}}$ and $m_{b\bar b\ell^-\ell^+}$, whose behavior is similar to that in benchmark point (i). After selection cuts, we have the total cross sections for both signal and background processes as
\begin{equation}
\sigma_{\mathrm{sig}}=1.13\kappa~(\mathrm{fb}),\quad\quad\mathrm{and}\quad\quad\sigma_{\mathrm{bkg}}=(17\xi+7.07)\times10^{-3}~(\mathrm{fb}).
\end{equation}

More details are shown in the appendices. For the other six benchmark points, we show the normalized distributions of $m_{b\bar b}$, $m_{b\bar b\ell^-\ell^+}$, and $m\mkern-10.5mu/$ in Appendix \ref{app:reco}. Following these distributions, we set the selection cuts on $m_{b\bar b}$ and $m_{b\bar b\ell^-\ell^+}$ for all the eight benchmark points. The selection cuts on $m_{b\bar b}$ are the same for all the eight benchmark points,
$100~\mathrm{GeV}\le m_{b\bar b}\le140~\mathrm{GeV}$; while the selection cuts on $m_{b\bar b\ell^-\ell^+}$ are sensitive to $m_{h_2}$, as shown in Table~\ref{tab:selection_cuts}. For all the eight benchmark points, $m\mkern-10.5mu/$ cannot be used to separate signal and background processes further after setting selection cuts on $m_{b\bar{b}}$ and $m_{b\bar b\ell^-\ell^+}$, thus it is only used as VBF tagging.
\begin{table}[ht]
\centering
\renewcommand{\arraystretch}{1.3}
\begin{tabular}{ccccc}
\hline\hline
$m_{h_2}$ [GeV]&700&1000&700&1000\\
$\sqrt{s}$ [TeV]&3&3&10&10\\
$m_{b\bar b\ell^-\ell^+}$ Cut [GeV]&$640-720$&$920-1020$&$640-720$&$920-1020$\\
\hline
$m_{h_2}$ [GeV]&1500&2000&3000&4500\\
$\sqrt{s}$ [TeV]&10&10&10&10\\
$m_{b\bar b\ell^-\ell^+}$ Cut [GeV]&$1300-1600$&$1700-2200$&$2400-3400$&$3400-5400$\\
\hline\hline
\end{tabular}
\caption{Selection cuts on $m_{b\bar b\ell^-\ell^+}$ for all the eight benchmark points.}
\label{tab:selection_cuts}
\end{table}

Following all the analysis procedure above, we show the cross sections for both signal and background processes for all the eight benchmark points, after each step (initial production, pre-selection cuts, and selection cuts), in Appendix \ref{app:cuts}. For background processes, we show both $\sigma_{\mathrm{b1-b4}}$ separately and the total cross section $\sigma_{\mathrm{bkg}}$. If $(c_2,c_{12})\le0.2$, we have $\xi\simeq\frac{c_2^4}{3c_2^2+2c_{12}^2}\lesssim1.3\times10^{-2}$, and thus for each benchmark point, the dominant background after selection cuts should be b2.

\section{Results}
\label{sec:results}

\begin{figure}[h]
\centering
\includegraphics[width=0.45\linewidth]{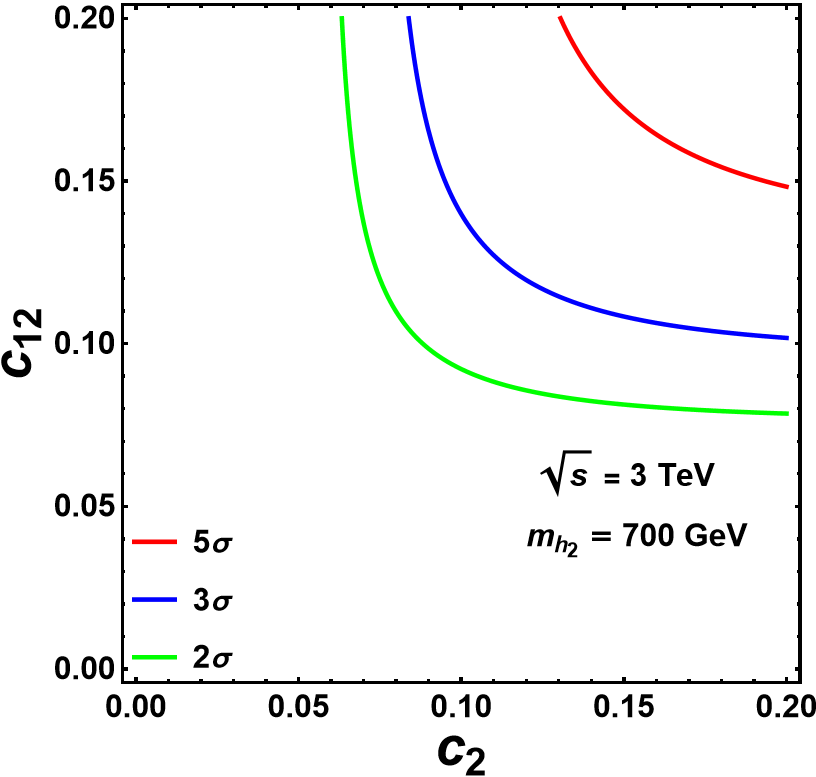}
\hspace{+6mm}
\includegraphics[width=0.45\linewidth]{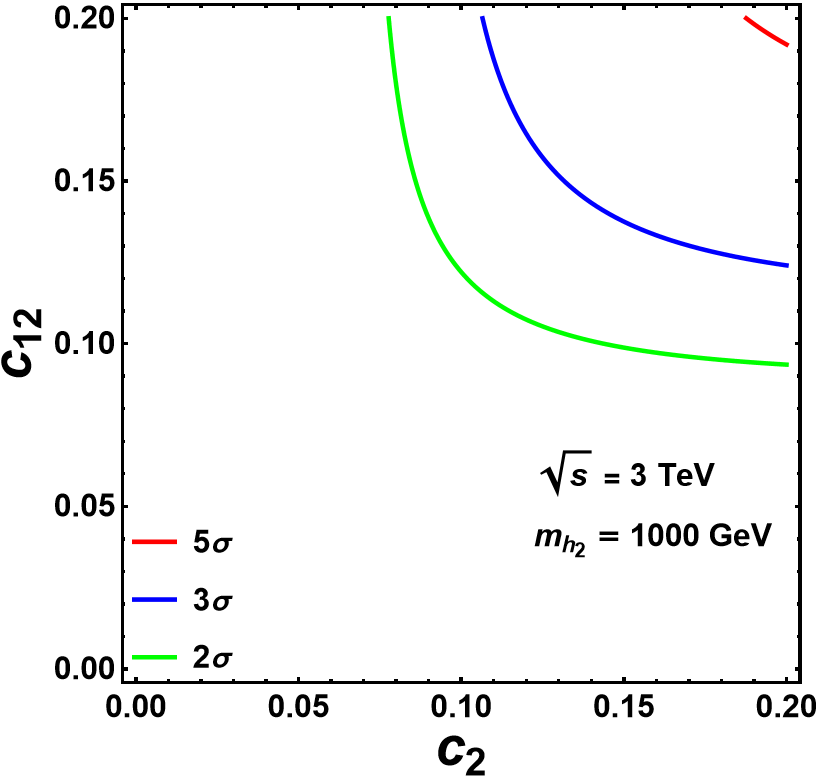}
\caption{Expected discovery potential for significance for points in $(c_2, c_{12})$-plane with $\sqrt{s}=3$~TeV and $L=0.9~\mathrm{ab}^{-1}$.
Benchmark points: $m_{h_2}\in\{700,1000\}$~GeV. The colored lines correspond to the boundaries with expected $2\sigma$ (green), $3\sigma$ (blue), and $5\sigma$ (red) significances, respectively.}
\label{fig:Result_3TeV}
\end{figure}

\begin{figure}[h]
\centering
\includegraphics[width=0.45\linewidth]{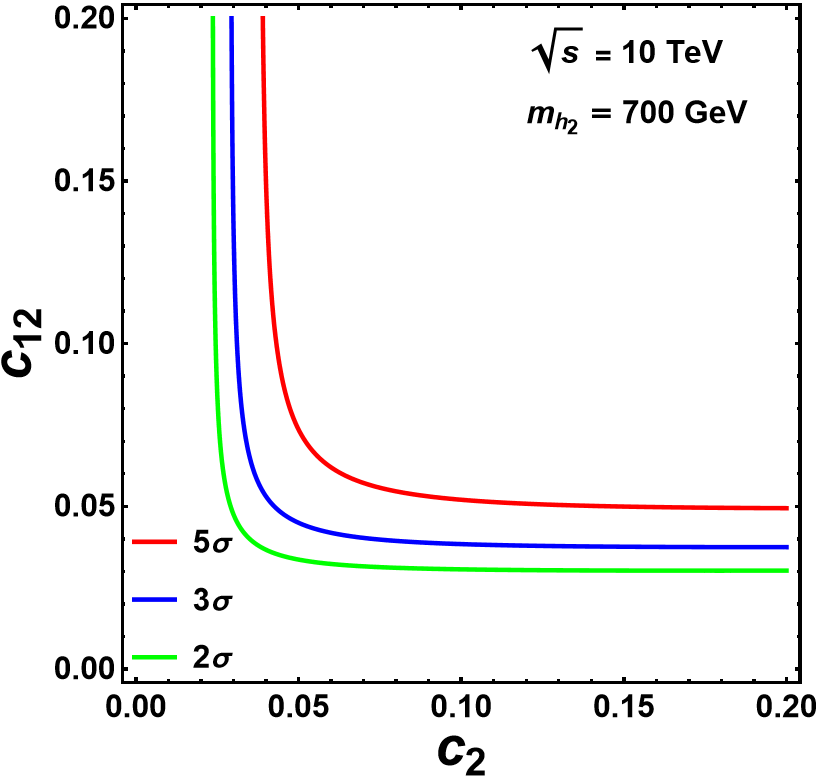}
\hspace{+6mm}
\includegraphics[width=0.45\linewidth]{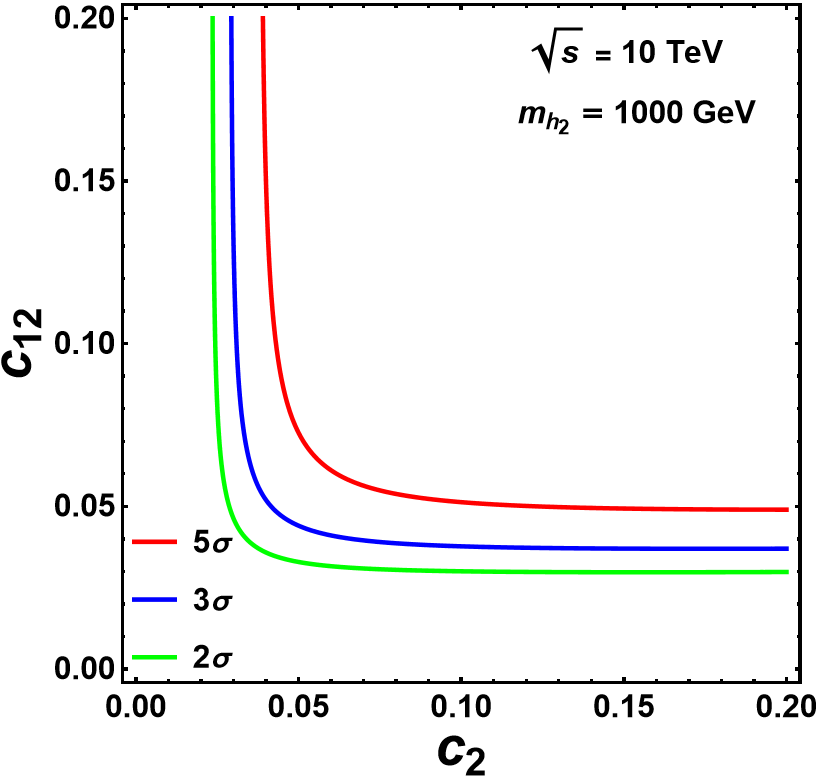}
\includegraphics[width=0.45\linewidth]{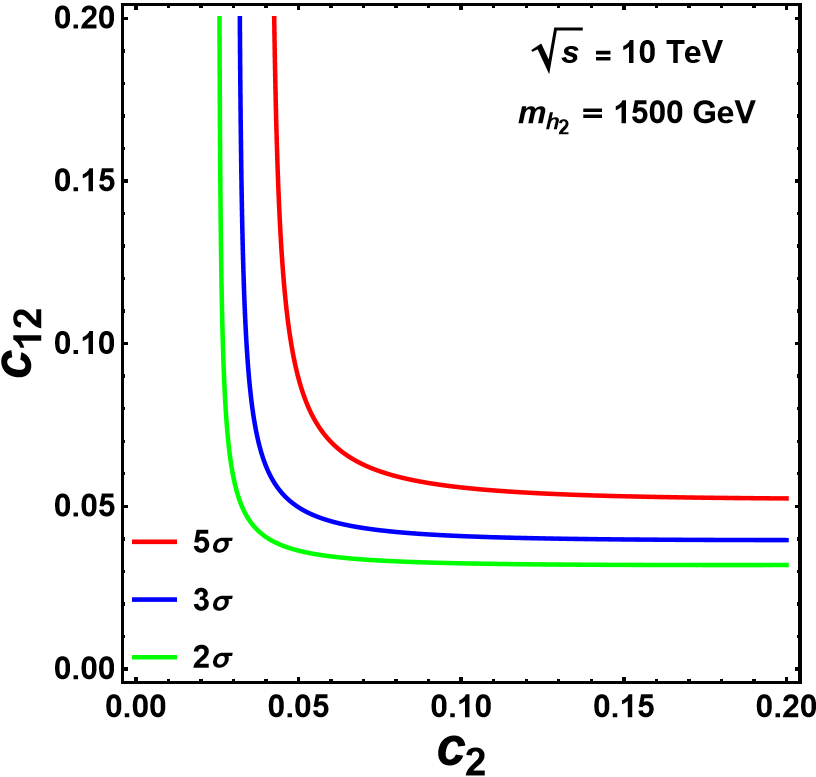}
\hspace{+6mm}
\includegraphics[width=0.45\linewidth]{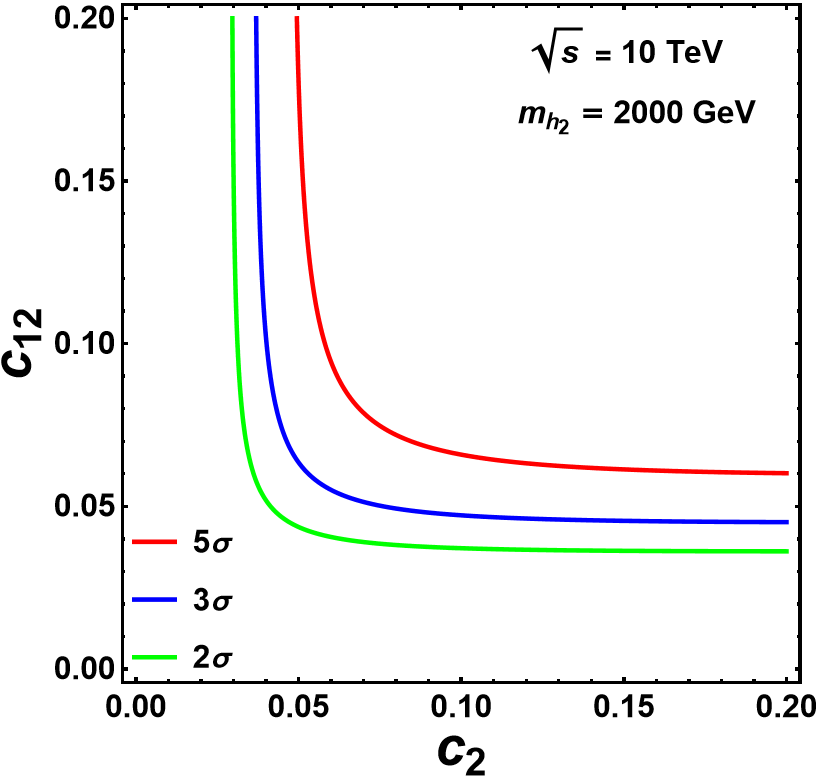}
\includegraphics[width=0.45\linewidth]{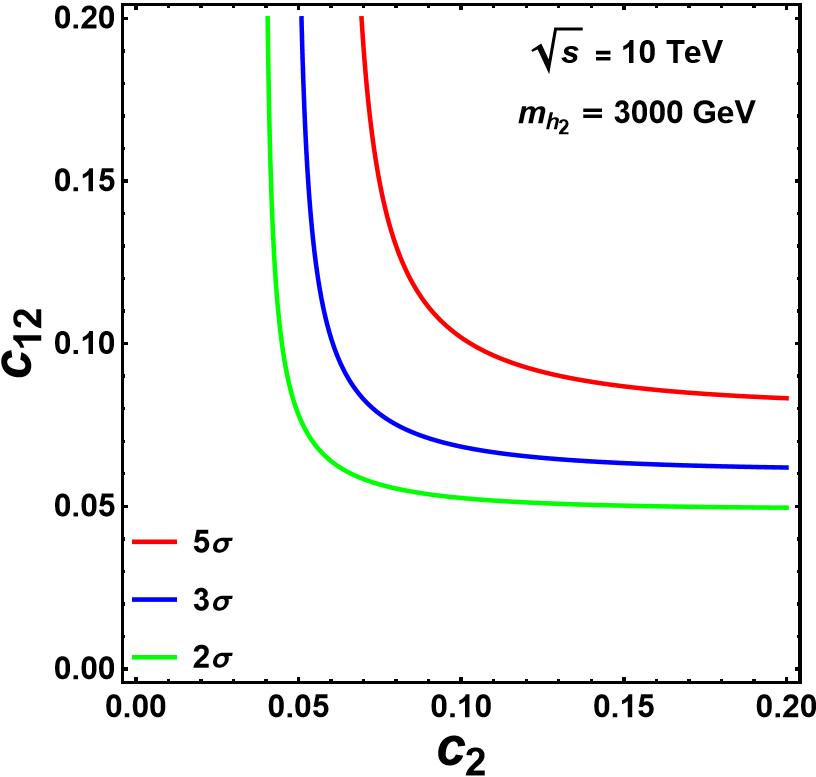}
\hspace{+6mm}
\includegraphics[width=0.45\linewidth]{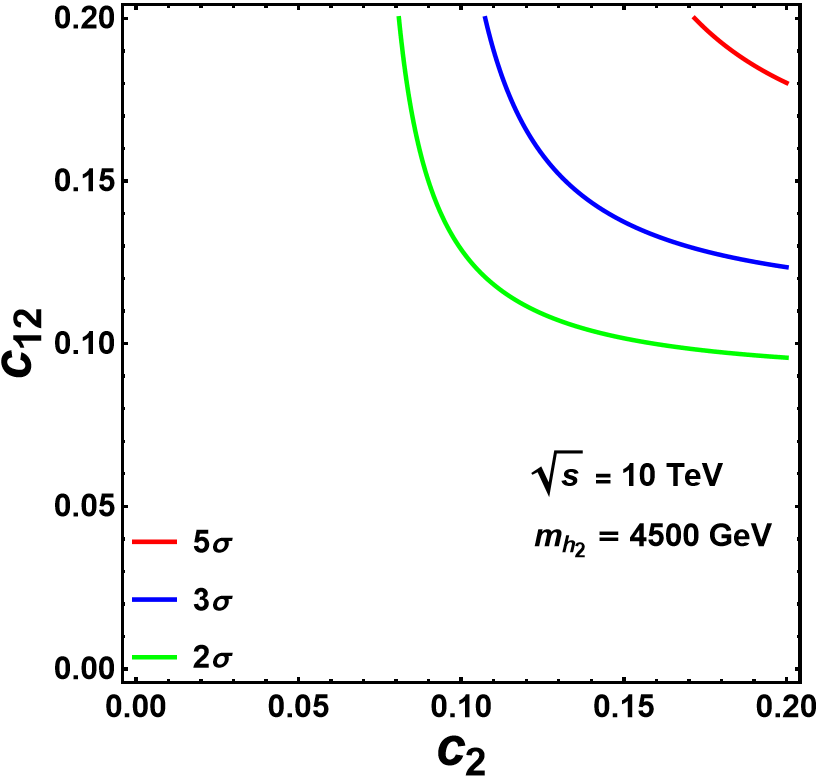}
\caption{Expected discovery potential for significance for points in $(c_2, c_{12})$-plane with $\sqrt{s}=10$~TeV and $L=10~\mathrm{ab}^{-1}$.
Benchmark points: $m_{h_2}\in\{700,1000,1500,2000,3000,4500\}$~GeV. The colored lines correspond to the boundaries with expected $2\sigma$ (green), $3\sigma$ (blue), and $5\sigma$ (red) significances, respectively.}
\label{fig:Result_10TeV}
\end{figure}

The expected event numbers of signal and background processes after selection cuts are
\begin{equation}
N_{\mathrm{sig}}=\sigma_{\mathrm{sig}}L,\quad\quad\mathrm{and}\quad\quad
N_{\mathrm{bkg}}=\sigma_{\mathrm{bkg}}L,
\end{equation}
where $L$ is the integrated luminosity.
The statistical significance is quantified through
\begin{equation}
\label{eq:significance}
\sigma_{\mathrm{stat}}=\sqrt{2\left[(N_{\mathrm{sig}}+N_{\mathrm{bkg}})\ln\left(1+\frac{N_{\mathrm{sig}}}{N_{\mathrm{bkg}}}\right)-N_{\mathrm{sig}}\right]},
\end{equation}
which increases as $L^{1/2}$.
Based on the analysis above, we present the expected discovery potential for CP violation in scalar sector
at muon colliders with $\sqrt{s}=3~(10)~\mathrm{TeV}$ and the corresponding integrated luminosity $L=0.9~(10)~\mathrm{ab}^{-1}$,
in Figs. \ref{fig:Result_3TeV} -- \ref{fig:Result_10TeV}.

For MuC with $\sqrt{s}=3~\mathrm{TeV}$ and corresponding integrated luminosity $L=0.9~\mathrm{ab}^{-1}$, assuming both $(c_2,c_{12})\lesssim0.2$,
CP violation in the scalar sector can be discovered at $5\sigma$ level up to $m_{h_2}\simeq1~\mathrm{TeV}$. For larger $m_{h_2}$, the cross section decreases quickly which leads to the weaker discovery potential. While for MuC with $\sqrt{s}=10~\mathrm{TeV}$ and corresponding integrated luminosity $L=10~\mathrm{ab}^{-1}$, assuming both $(c_2,c_{12})\lesssim0.2$, CP violation in the scalar sector can be discovered at $5\sigma$ level up to $m_{h_2}\simeq4.5~\mathrm{TeV}$, because of the larger cross section and integrated luminosity. When $m_{h_2}\lesssim3~\mathrm{TeV}$, the discovery potential changes slowly, and the case with parameters $(c_2,c_{12})\simeq0.1$ can be discovered at $5\sigma$ level. The results show significant advantages for future MuC to test CP violation in the scalar sector, especially for the heavy $h_2$ scenario.

\section{Conclusions and discussions}
\label{sec:conclusion}

We propose a model-independent strategy to test CP violation in an extended scalar sector at future muon colliders (MuC). The key observation is  the existence of the single process
\begin{equation}
V V \to h_2 \to Z h_1, \quad\quad (\mathrm{with}~V=W^\pm,Z).
\end{equation}
It already requires both couplings $c_2\neq 0$ and $c_{12}\neq 0$. Therefore, an observation of this channel is sufficient to establish CP violation in the scalar sector within the effective framework of Eq.~(\ref{eq:effint}).
The absorbers installed to mitigate beam-induced backgrounds (BIB)~\cite{Mokhov:2011zzd,Mokhov:2014hza,DiBenedetto:2018cpy,Bartosik:2019dzq} imply that the signal receives contributions from both $W^+W^-$ and $ZZ$ fusion, because the very forward/backward $\mu^{\pm}$ cannot be detected and thus behave as missing energy.

In our collider study, we focus on the clean final state $Z(\to\ell^+\ell^-)\,h_1(\to b\bar b)$, exploit the reconstructed masses $m_{b\bar b\ell^-\ell^+}$ and $m_{b\bar b}$, and use simple, robust selections to suppress backgrounds. After selection cuts, the dominant background is b2 for all benchmark points, while b1 is efficiently suppressed by $m_{b\bar{b}}$ cut, the smallness of $\xi$ and $\mathrm{Br}_{Z\to b\bar{b}}$. The sensitivity trends obtained at $\sqrt{s}=3$ and $10~\mathrm{TeV}$ (with the respective integrated luminosities $L=0.9$ and $10~\mathrm{ab}^{-1}$ used in Sec.~\ref{sec:simulation}) show that the VBF topology offers competitive discovery reach for a wide range of $m_{h_2}\sim\mathcal{O}(\mathrm{TeV})$ at future MuC. Especially, for the $\sqrt{s}=10$~TeV future MuC with $L=10~\mathrm{ab}^{-1}$, $(c_{2},c_{12})\simeq0.1~(0.2)$ can be discovered at $5\sigma$ level up to a rather heavy $m_{h_2}\lesssim3~(4.5)~\mathrm{TeV}$.

Our analysis targets the multi-TeV regime where VBF processes naturally benefit from the logarithmic enhancement of gauge boson radiation, making MuC an efficient ``gauge boson collider''. The considered MuC parameters $\sqrt{s}$ and $L$ are consistent with current community roadmaps and interim reports for a staged MuC program~\cite{InternationalMuonCollider:2024jyv,USMCC:2025WP}. Within this context, the $V V\to h_2\to Z h_1$ channel provides a direct handle on CP violation that is complementary to EDM constraints and to Yukawa-sector probes discussed in Sec.~\ref{sec:intro}. Importantly, our selection is compatible with realistic machine--detector interface conditions at MuC, where forward/backward shielding against BIB \cite{Mokhov:2011zzd,Mokhov:2014hza,DiBenedetto:2018cpy,Bartosik:2019dzq} motivates restricted angular acceptance in the final states.

In this work, we focus on a general $h_2$-interaction in Eq.~(\ref{eq:effint}), without assuming a UV complete model, targeting this special type of CP-violation in the heavy-scalar sector. The main limitation of this method is that we must first discover the additional scalar $h_2$, and it must interact with the $Z$-boson and SM-like Higgs boson $h_1$ at tree level, following Eq.~(\ref{eq:effint}). Once these conditions are satisfied, this method is universal for BSM models which contain one or more heavy scalar(s) with such type of interactions. As mentioned in Ref. \cite{Li:2016zzh}, some CP-conserving neutral state $h_2$ may also interact with $VV$ and $Zh_1$ at loop level. However, such scenarios are distinguishable from the case we discuss in this paper, because loop-induced effective interactions always lead to small enough cross sections, corresponding to an effective $(c_{2},c_{12})\lesssim\mathcal{O}(10^{-2})$. That means our method at MuC is sufficient to confirm CP violation in the scalar sector.

In this work, we did not consider $h_2$-interactions beyond Eq.~(\ref{eq:effint}). In a specific UV complete model, $h_2$ of course may interact with other SM particles (depending strongly on the concrete model construction), which may lead to rich phenomenology. For example, as discussed in Sec. \ref{sec:intro}, if $h_2$ interact with fermions, it may also lead to CP violation effects in Yukawa sector. Since we focus on the specific type of CP violation in this paper in a model-independent way without further assumptions on the UV complete model, here we do not discuss other CP violation effects beyond the interaction (\ref{eq:effint}), and we leave such studies to some forthcoming projects. If $h_2$ has the interactions beyond Eq.~(\ref{eq:effint}), there will also be additional decay channels of $h_2$, which will modify the total width of $h_2$ in Eq.~(\ref{eq:totalwidth}) and hence the decay branching ratio $\mathrm{Br}_{h_2\to Zh_1}$. It will not break the effectiveness of the method, but modifies the detailed discovery potential shown in Sec. \ref{sec:results} a bit.

In summary, observing $V V\to h_2\to Z h_1$ with the selections outlined in this work would constitute compelling evidence for CP violation in the scalar sector; conversely, the non-observation can be translated into robust, model-independent bounds on the coupling combination $(c_2, c_{12})$, providing a clear target for future MuC programs.


\appendix

\section{Sum rule for the couplings $c_1,~c_2$, and $c_{12}$ in a simplified scenario}
\label{app:sumrule}
Generally, for a scalar multiplet $\phi_I$ with the weak isospin $I$ and weak hypercharge $Y$, the electroweak covariant derivative can be written as
\begin{equation}
D_\mu\phi_I\equiv\left(\partial_\mu-\textrm{i}gW^a_\mu\tau^a_I-\textrm{i}g'\frac{Y}{2}B_\mu\right)\phi_I,
\end{equation}
where $g,g'$ are gauge coupling constants, $W_\mu^a$ and $B_\mu$ are gauge fields of $\mathrm{SU}(2)_L$ and $\mathrm{U}(1)_Y$ groups separately. $\phi_I$ forms a $2I+1$-plet of $\mathrm{SU}(2)_L$ group, which must be a complex field if $Y\neq0$. $\tau^a_I$ are three generators of $\mathrm{SU}(2)_L$ group in the representation of dimension $2I+1$; for $I=1/2$, $\tau^a_I$ become the Pauli matrices. We usually choose $\tau_I^3$ diagonal as
\begin{equation}
\tau_I^3=\mathrm{diag}\{I,I-1,\ldots,1-I,-I\}.
\end{equation}
For a component of $\phi_I$, we denote it as $\phi_{I_3}$ which contains the isospin third component $I_3$. The covariant derivative containing the neutral gauge fields can be written as
\begin{eqnarray}
D_\mu\phi_{I_3}&\supset&\left[\partial_\mu-\textrm{i}gI_3\left(s_WA_\mu+c_WZ_\mu\right)-\textrm{i}g'\frac{Y}{2}\left(c_WA_\mu-s_WZ_\mu\right)\right]\phi_{I_3}\nonumber\\
\label{eq:code}
&=&\left[\partial_\mu-\textrm{i}|e|\left(I_3+\frac{Y}{2}\right)A_\mu-\textrm{i}\frac{g}{c_W}\left(I_3c_W^2-\frac{Y}{2}s^2_W\right)Z_\mu\right]\phi_{I_3}.
\end{eqnarray}
Here $s_W(c_W)\equiv\sin\theta_W(\cos\theta_W)$, and $A_\mu$ is the photon field. Based on Eq.~(\ref{eq:code}), we straightforwardly have the extension of Gell-Mann-Nishijima relation to electroweak interaction as $Q=I_3+Y/2$. For a neutral component $\phi_{I_3}$, we have $I_3=-Y/2$, and thus the covariant derivative~(\ref{eq:code}) is reduced to
\begin{equation}
D_\mu\phi_{I_3}\supset\left(\partial_\mu+\textrm{i}\frac{gY}{2c_W}Z_\mu\right)\phi_{I_3},
\end{equation}
depending on the weak hypercharge $Y$.

Assuming custodial symmetry in the Higgs sector, we restrict to the simplified case where the scalar vacuum expectation values (VEVs) appear only in $\mathrm{SU}(2)_L$ singlets or doublets \cite{ParticleDataGroup:2024cfk}.
Here we set the VEVs of all other types of Higgs multiplets zero for simplicity. For all the neutral singlets, since $I=Y=0$, they cannot lead to interactions with SM gauge bosons like in Eq.~(\ref{eq:effint}). In any UV complete model, no matter how many scalar doublets it contains, we can always perform a rotation among the doublets, to ensure only one of the doublets contains VEV; and we then denote it as $\phi_{\mathrm{SM}}$ which contains the dominant componenat of $h_1$. In a general model, the extended scalar sector may be very complex. However, we can consider a simplified scenario, in which only one extended scalar multiplet $\phi_I$ contributes to the interaction (\ref{eq:effint}) significantly, or equivalently, $h_2$ mainly comes from $\phi_I$. In this scenario, the neutral scalar sector contains
\begin{equation}
\label{eq:basis}
\phi_{\mathrm{SM}}\supset\frac{v_{\mathrm{SM}}+h}{\sqrt{2}},\quad\quad\mathrm{and}\quad\quad
\phi_I\supset\phi_{I_3}=\frac{H+\textrm{i}H'}{\sqrt{2}}.
\end{equation}
In the basis (\ref{eq:basis}), the $h$ component contributes only to $hVV$ coupling, while the $H$ and $H'$ components contribute only to the $HH'Z$ coupling, as
\begin{equation}
\mathcal{L} \supset \left(\frac{g^2 v}{2} W^{+\mu} W^-_\mu + \frac{g^2 v}{4 c_W^2} Z^{\mu} Z_\mu\right)h
+ \frac{g Y}{2 c_W} Z_\mu \left(H \partial^\mu H' - H' \partial^\mu H\right).
\end{equation}
The scalar mass eigenstates $h_{1,2,3}$ can be defined through an orthogonal matrix $\mathcal{R}$ (whose elements are defined as $\mathcal{R}_{ij}$) as
\begin{equation}
(h_1,h_2,h_3)=(h,H,H')\mathcal{R}^T.
\end{equation}
Comparing with Eq.~(\ref{eq:effint}), we straightforwardly have
\begin{equation}
c_1=\mathcal{R}_{11},\quad c_2=\mathcal{R}_{21},\quad c_{12}=Y(\mathcal{R}_{12}\mathcal{R}_{23}-\mathcal{R}_{13}\mathcal{R}_{22})=Y\mathcal{R}_{31}.
\end{equation}
$Y=0$ leads to a vanishing $c_{12}$; equivalently, $c_{12}\neq0$ requires $|Y|$ to be a positive integer. We then obtain the sum rule of the couplings as
\begin{equation}
c_1^2+c_2^2+\frac{c_{12}^2}{Y^2}=1.
\end{equation}
It means that the sum rule for the couplings $c_1,~c_2$, and $c_{12}$ is sensitive to the UV complete model. As an example, for 2HDM which requires $|Y|=1$, we simply have $c_1^2+c_2^2+c^2_{12}=1$. For any $|Y|\geq1$, the parameter region $c_2^2+c_{12}^2<1-c_1^2$ is still not excluded by experiments, and thus we can typically choose such benchmarks in the text.

\section{Additional reconstruction distributions}
\label{app:reco}

\begin{figure}[h]
\centering
\includegraphics[width=0.33\linewidth]{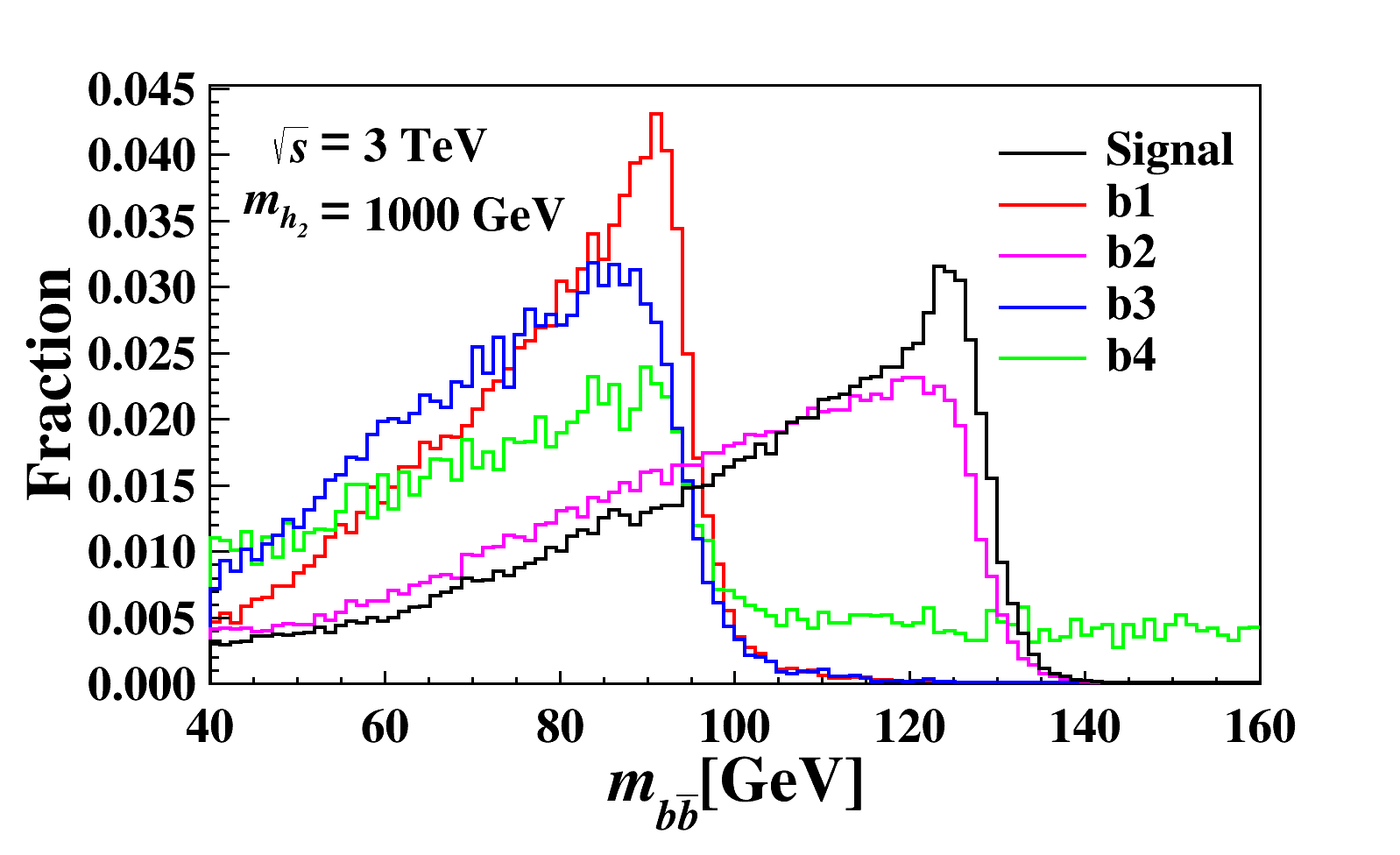}
\hspace{-6mm}
\includegraphics[width=0.33\linewidth]{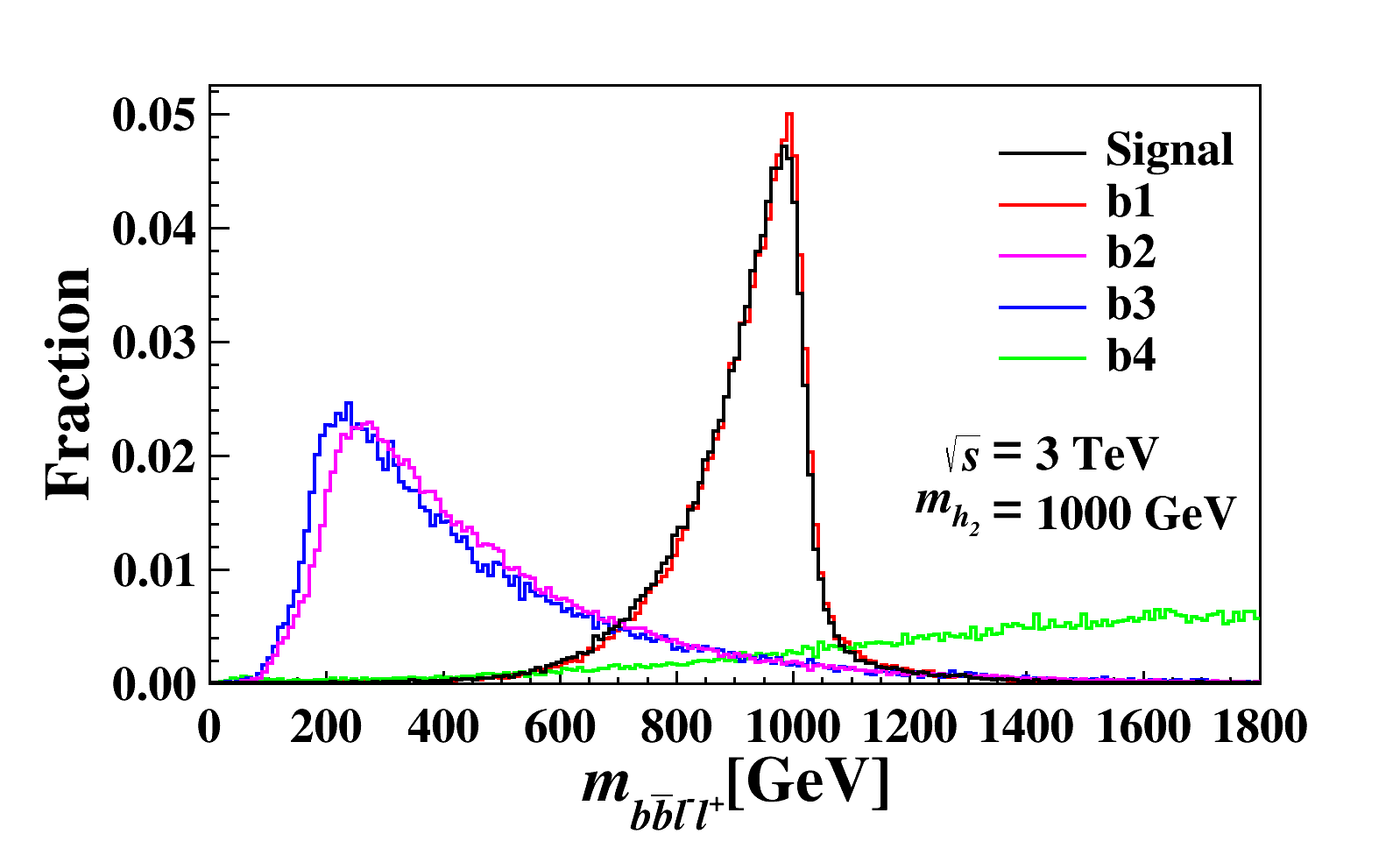}
\hspace{-6mm}
\includegraphics[width=0.33\linewidth]{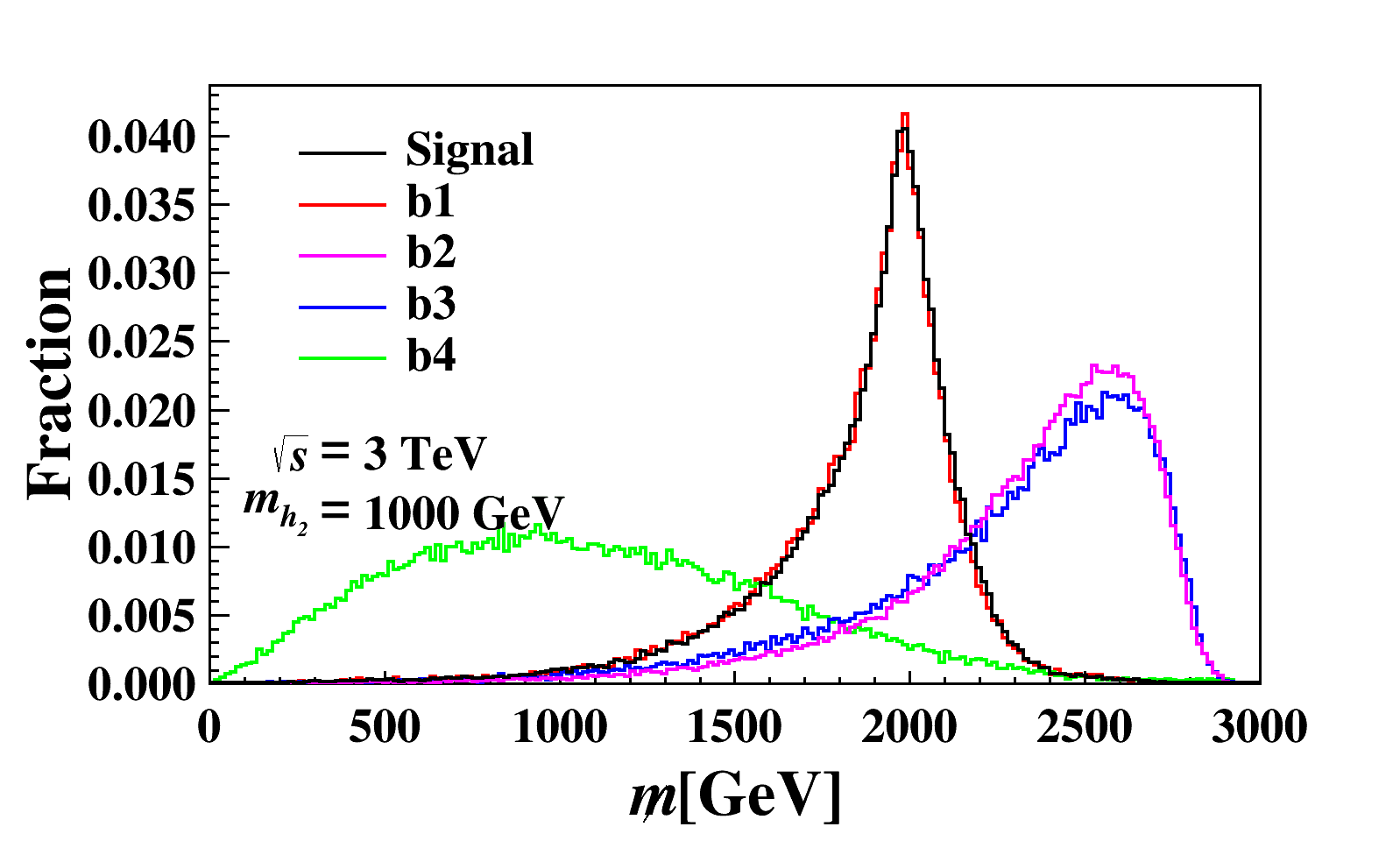}
\includegraphics[width=0.33\linewidth]{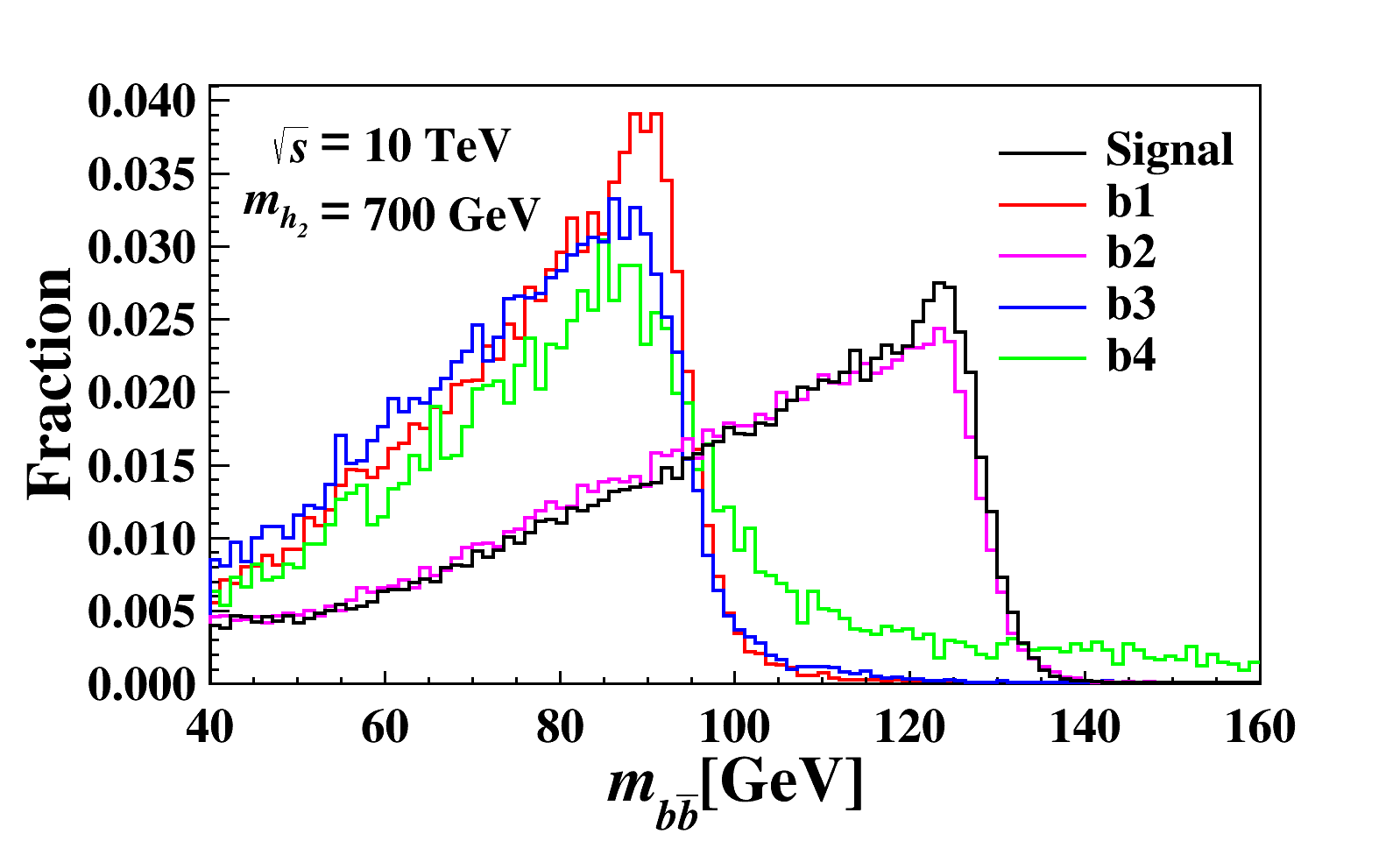}
\hspace{-6mm}
\includegraphics[width=0.33\linewidth]{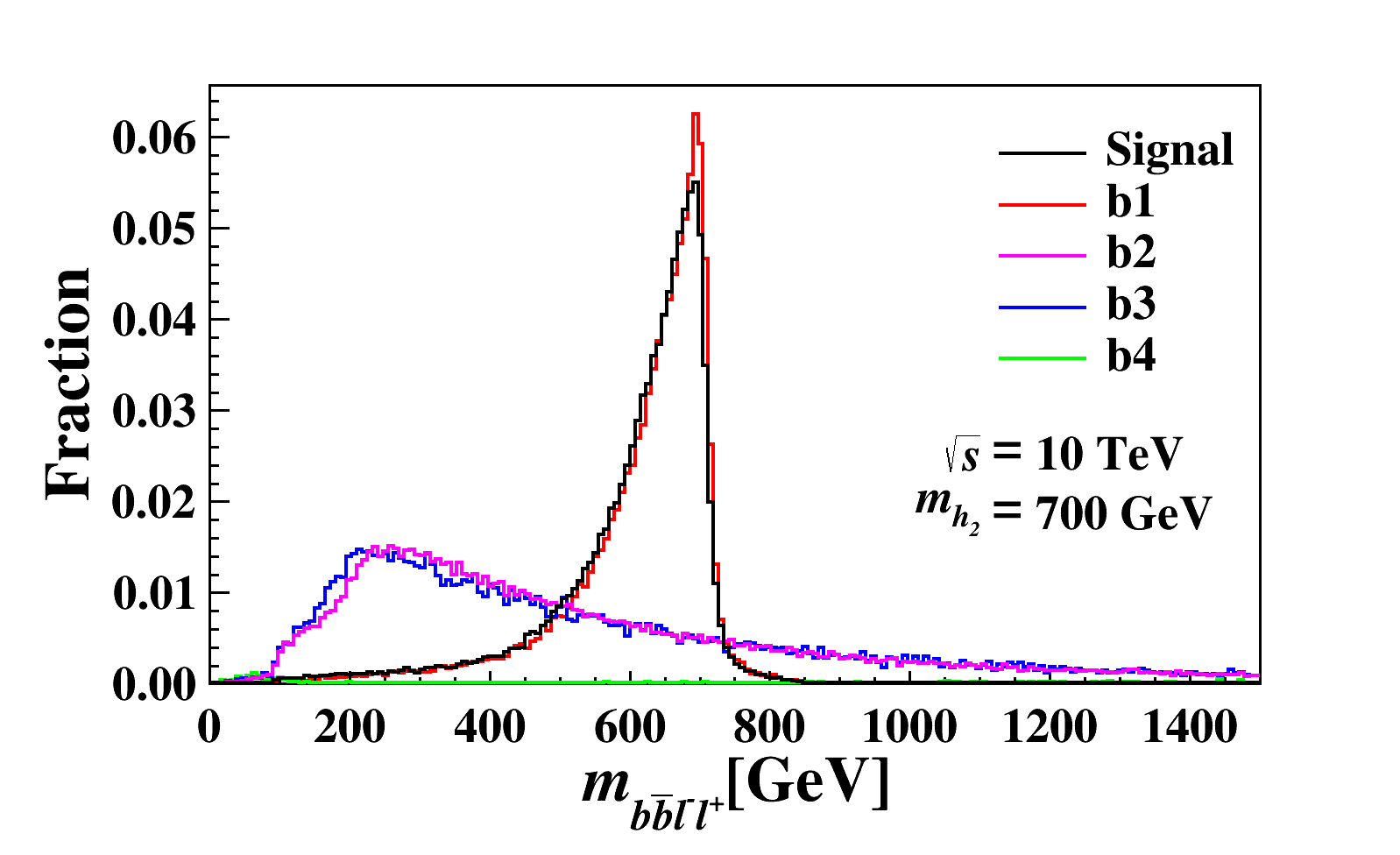}
\hspace{-6mm}
\includegraphics[width=0.33\linewidth]{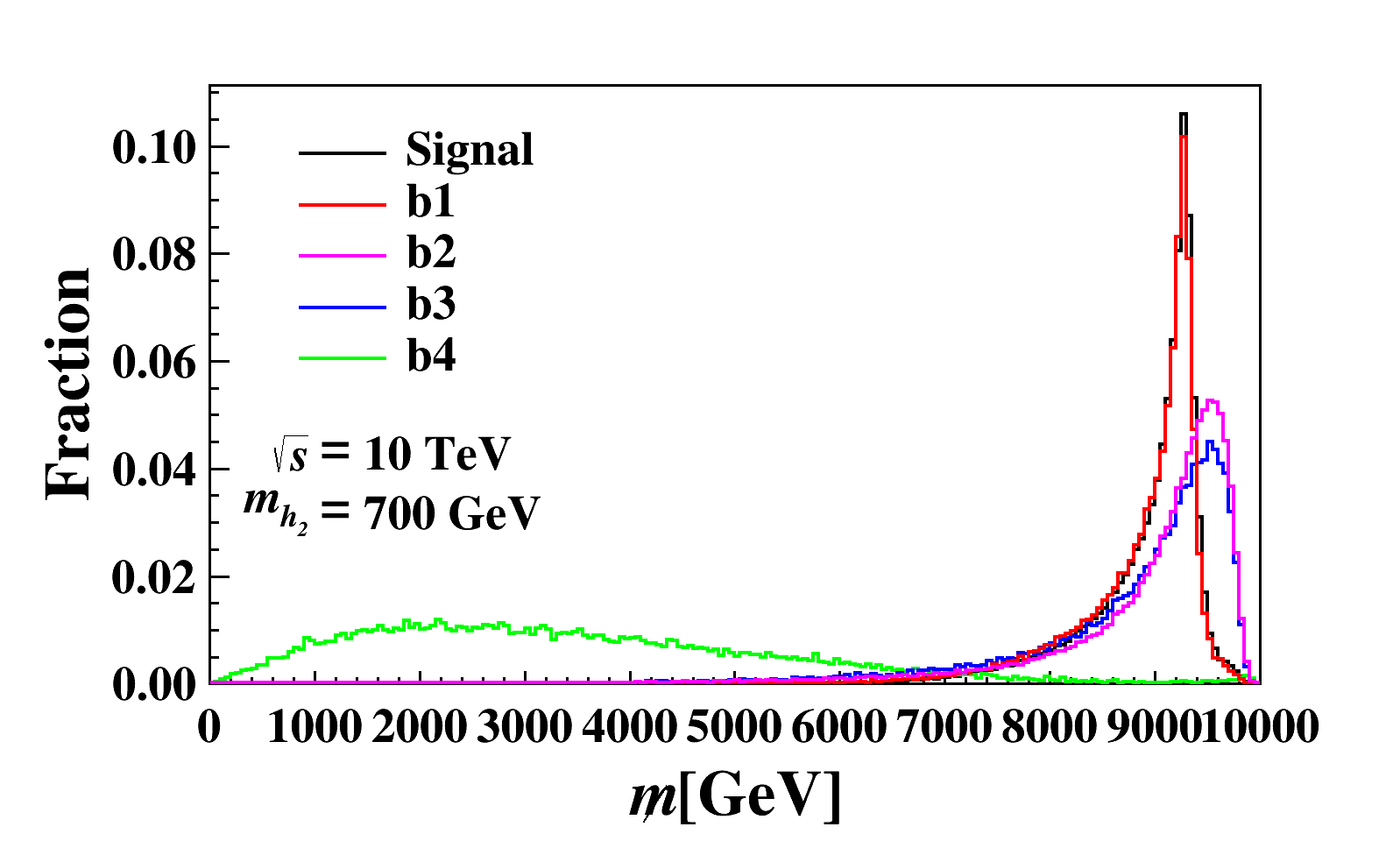}
\includegraphics[width=0.33\linewidth]{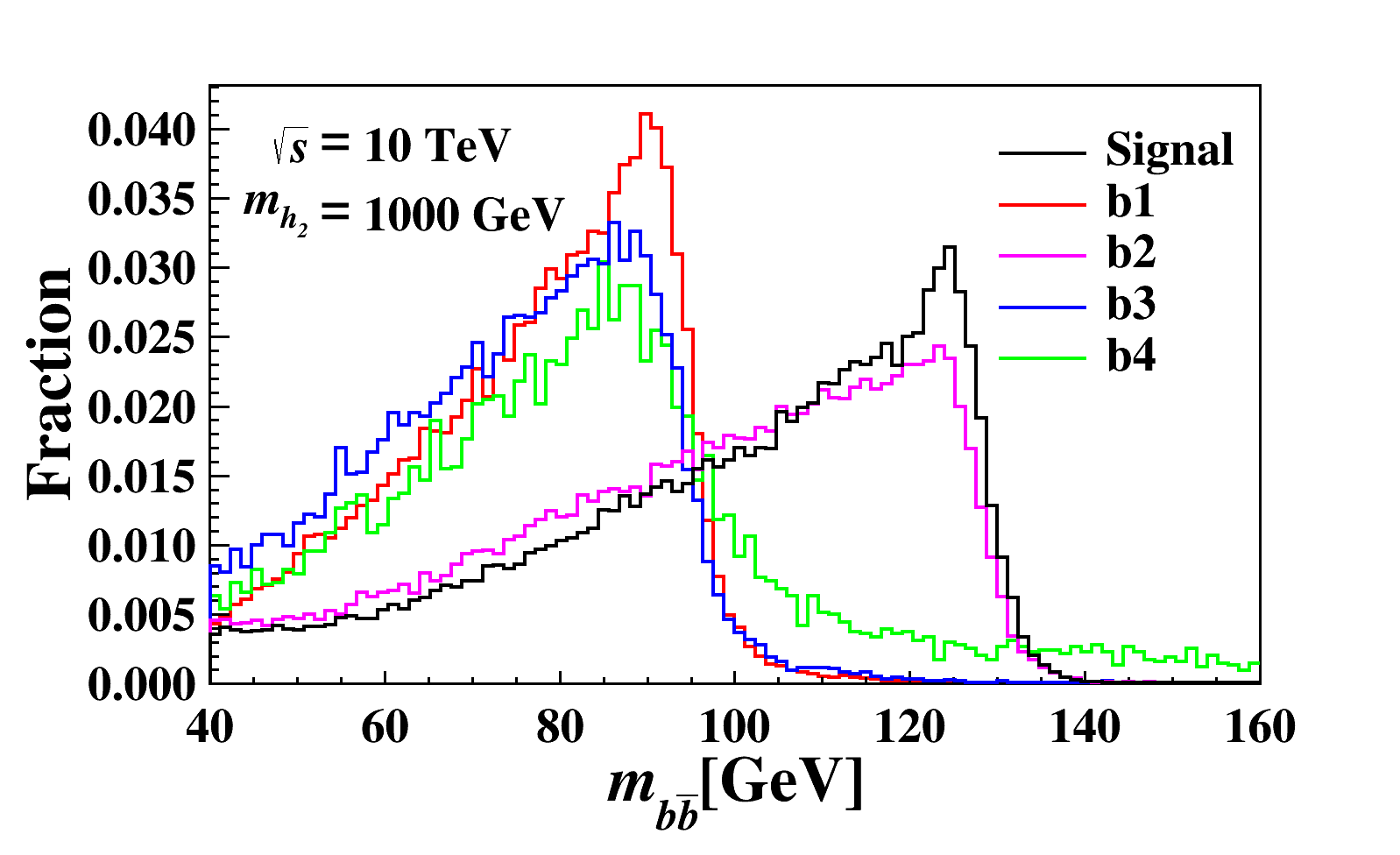}
\hspace{-6mm}
\includegraphics[width=0.33\linewidth]{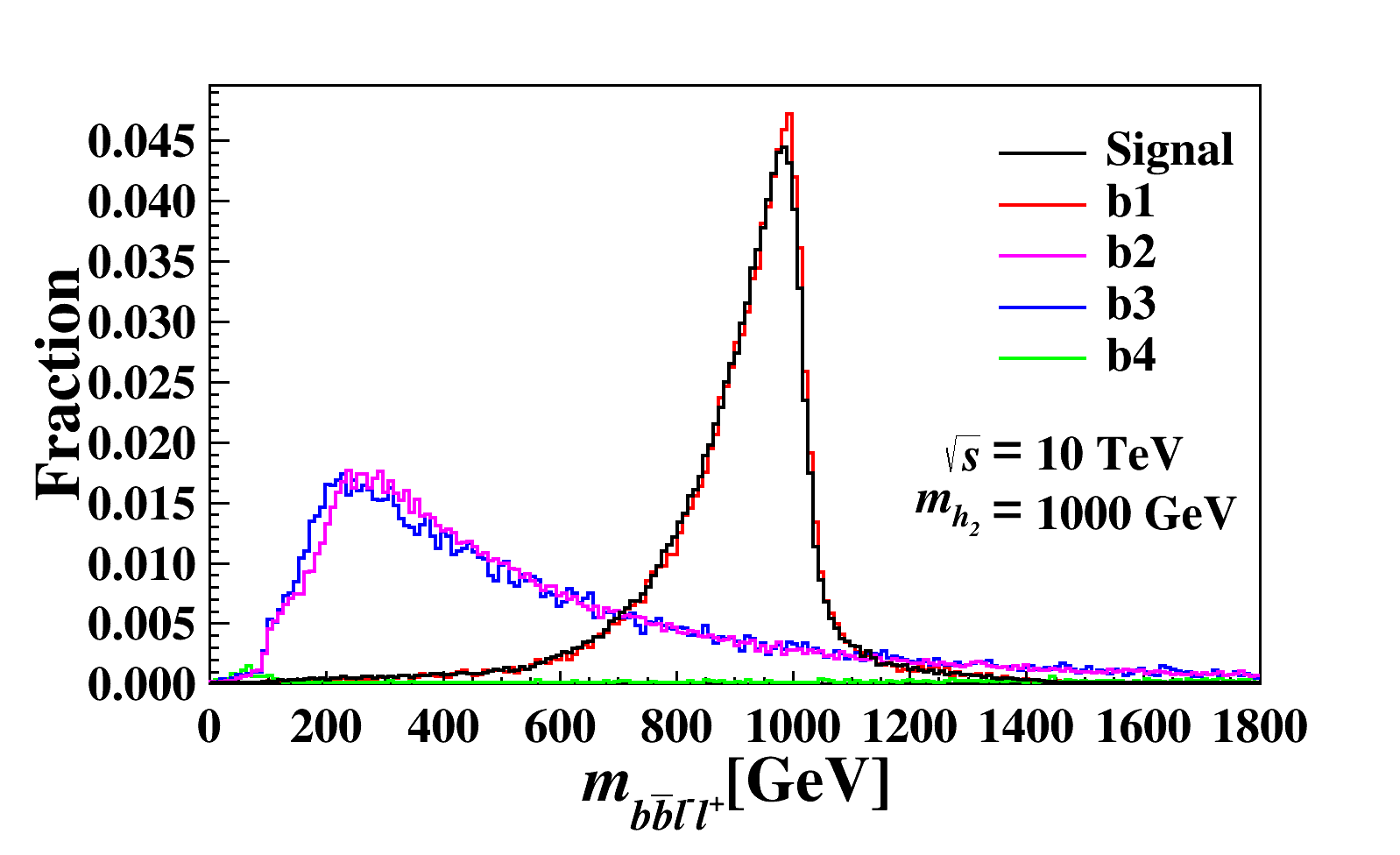}
\hspace{-6mm}
\includegraphics[width=0.33\linewidth]{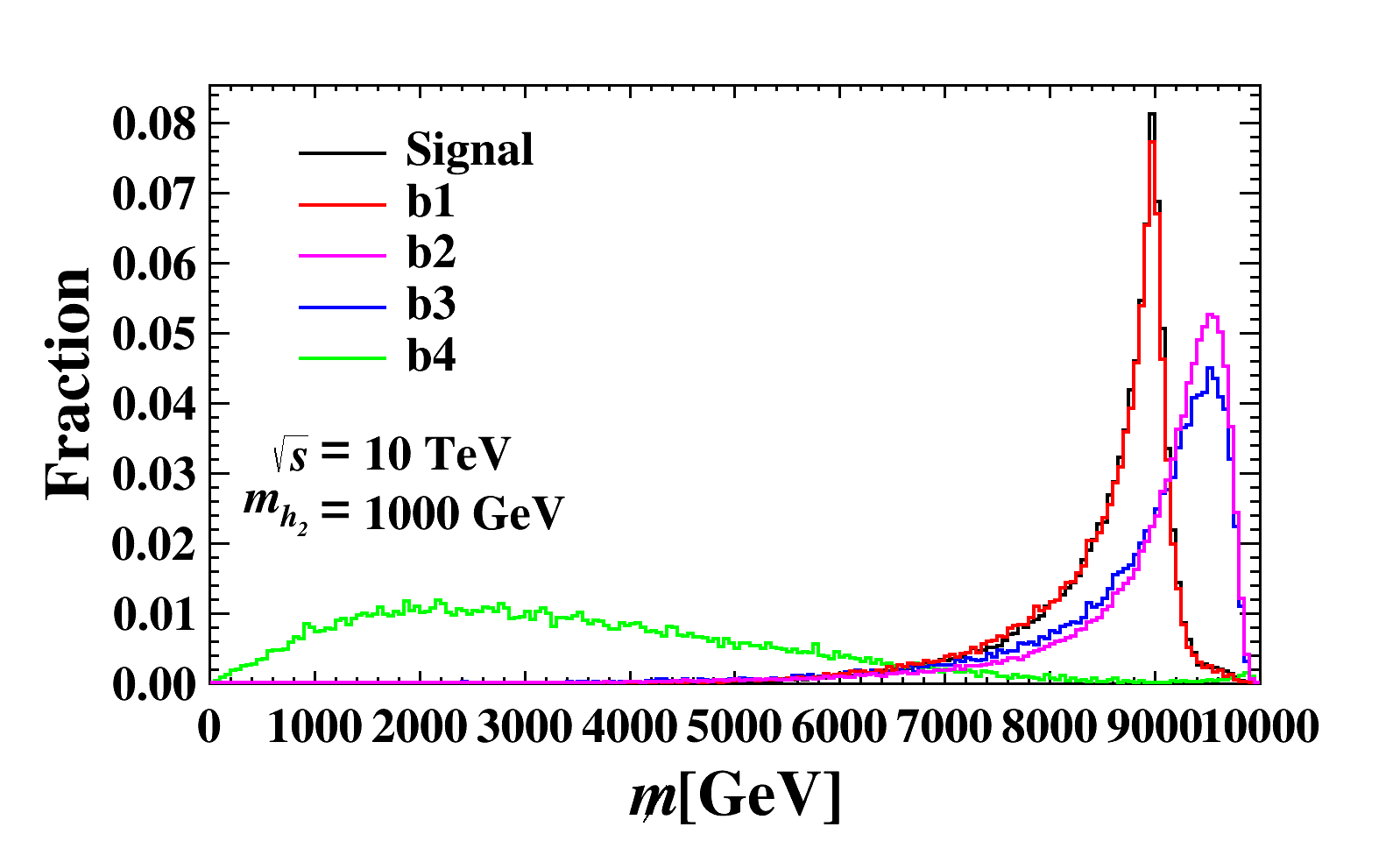}
\includegraphics[width=0.33\linewidth]{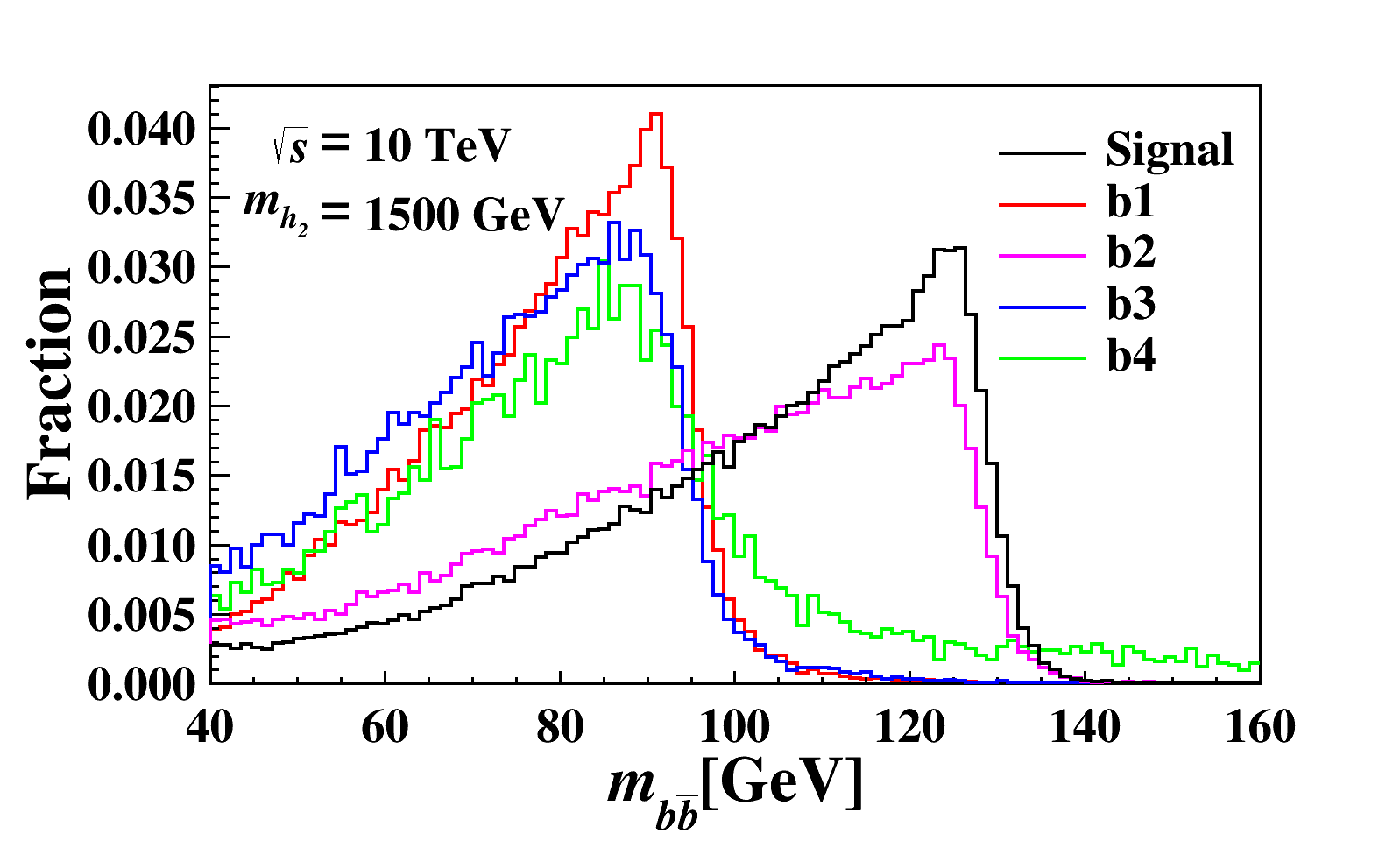}
\hspace{-6mm}
\includegraphics[width=0.33\linewidth]{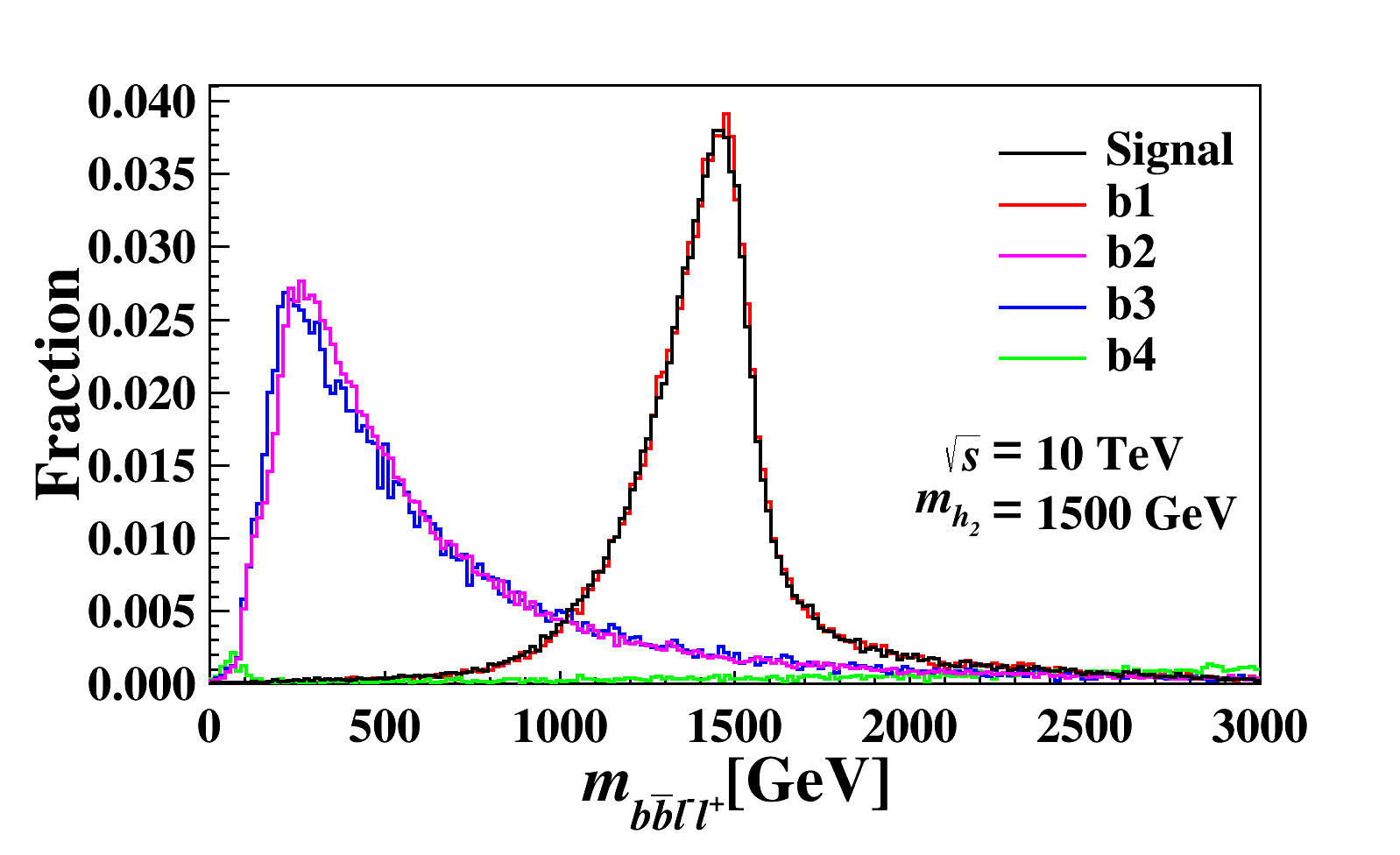}
\hspace{-6mm}
\includegraphics[width=0.33\linewidth]{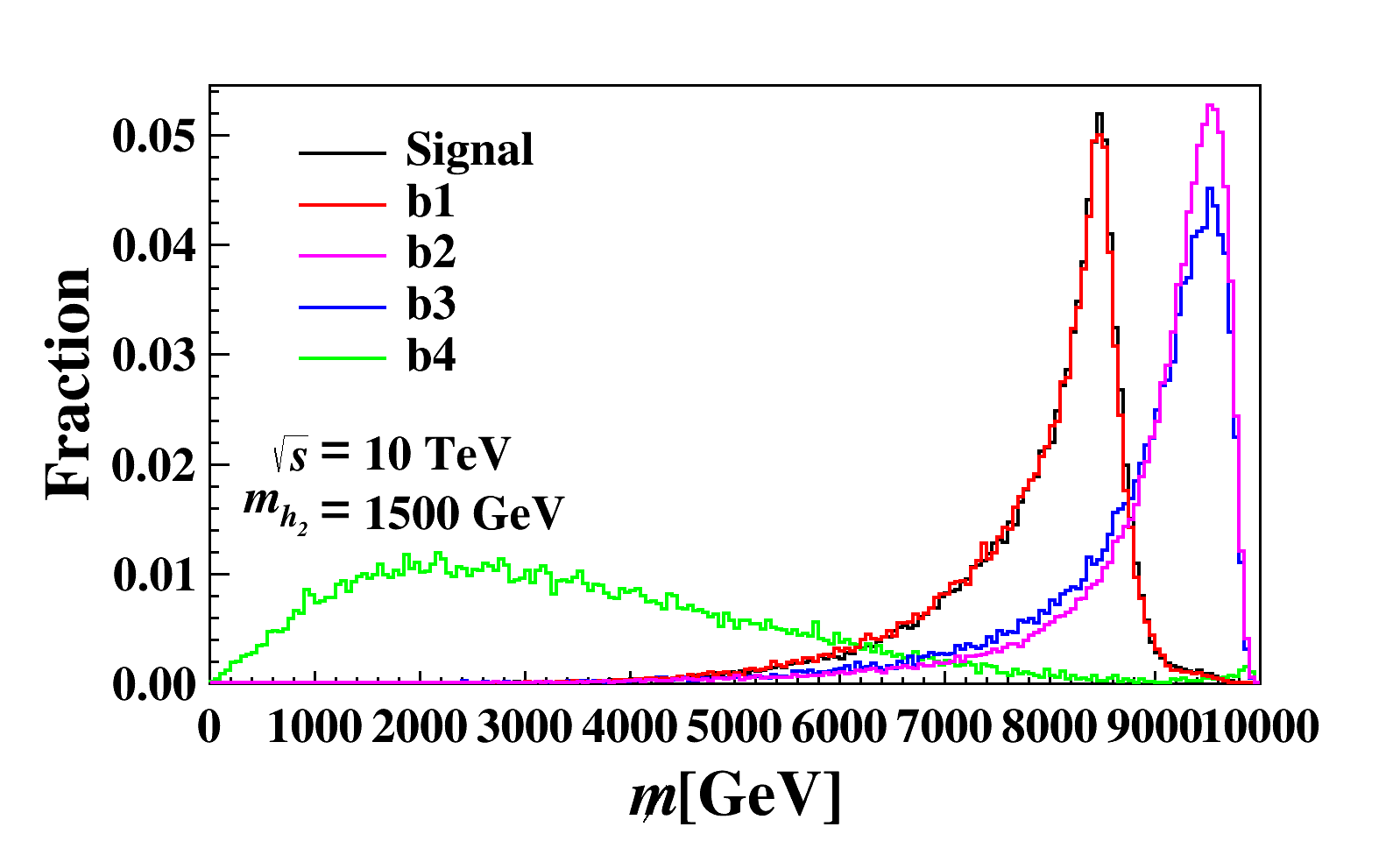}
\includegraphics[width=0.33\linewidth]{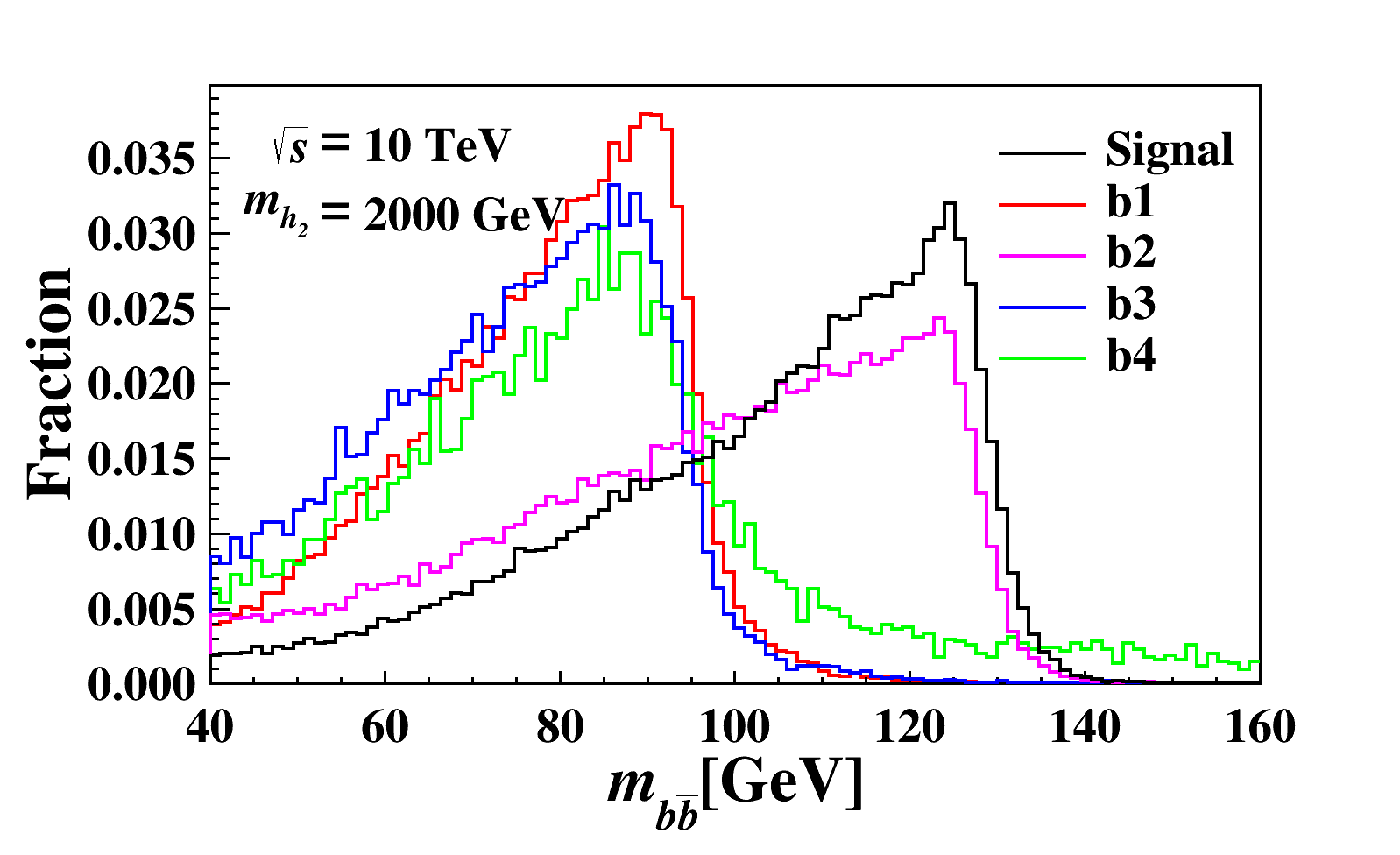}
\hspace{-6mm}
\includegraphics[width=0.33\linewidth]{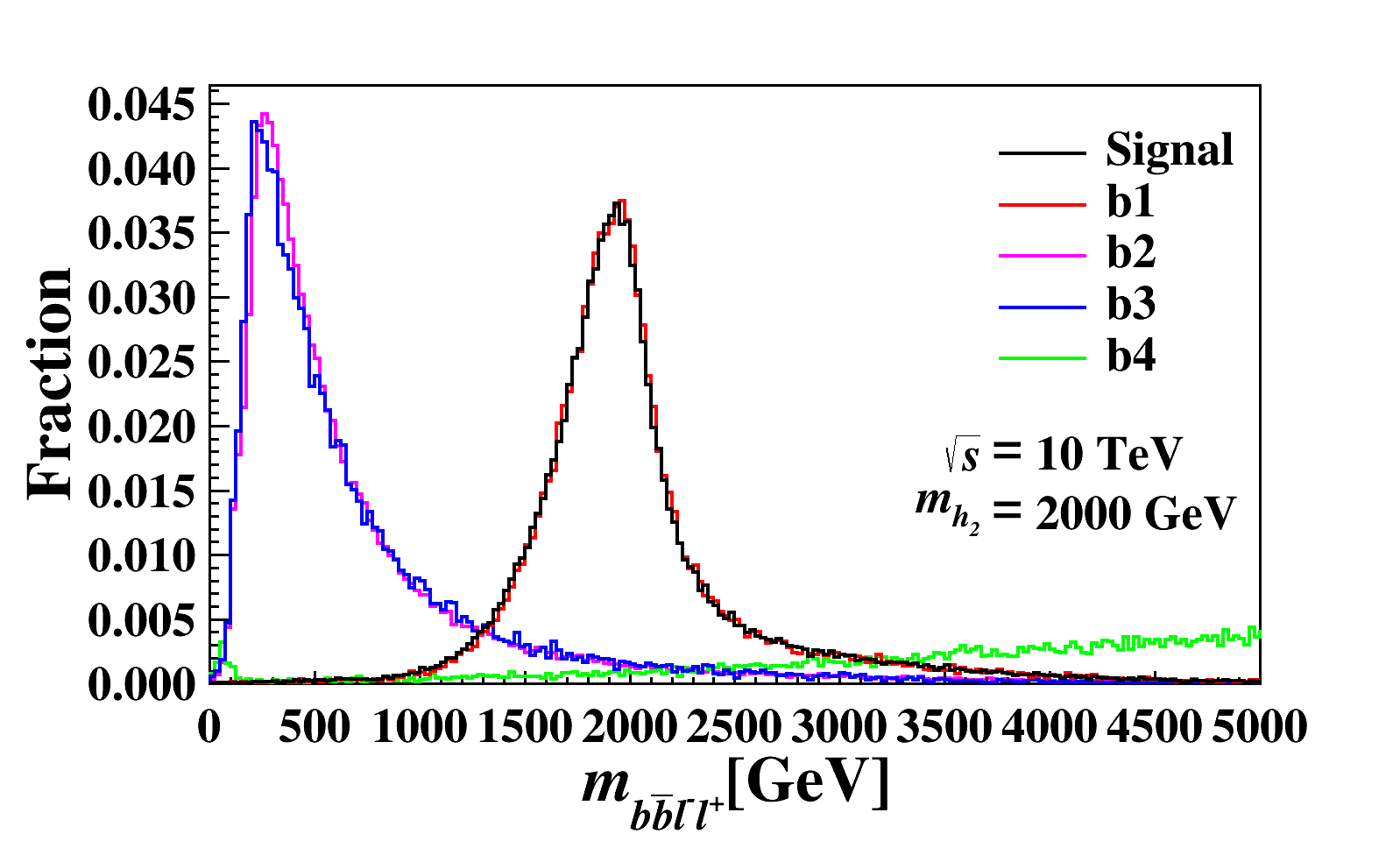}
\hspace{-6mm}
\includegraphics[width=0.33\linewidth]{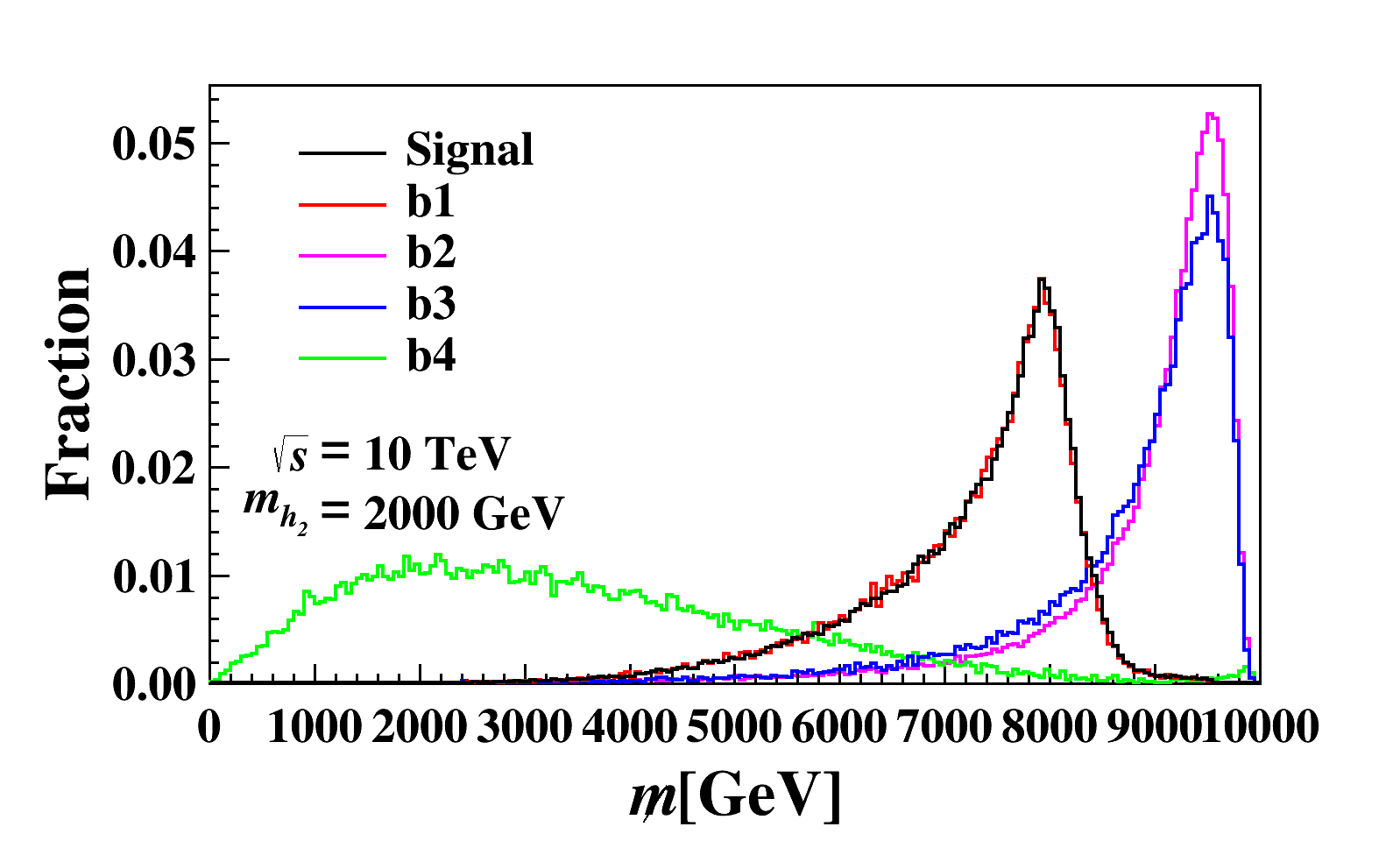}
\includegraphics[width=0.33\linewidth]{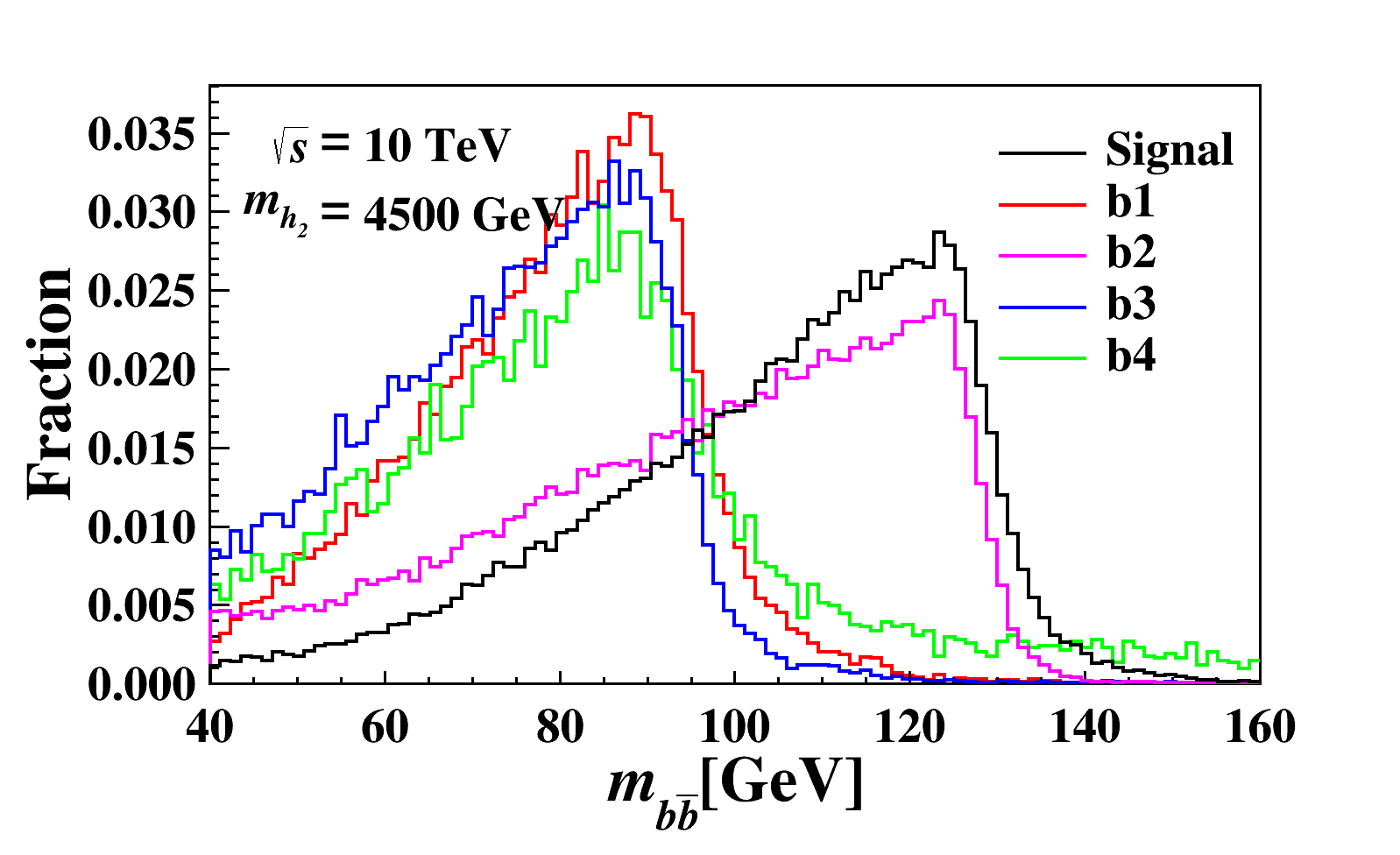}
\hspace{-6mm}
\includegraphics[width=0.33\linewidth]{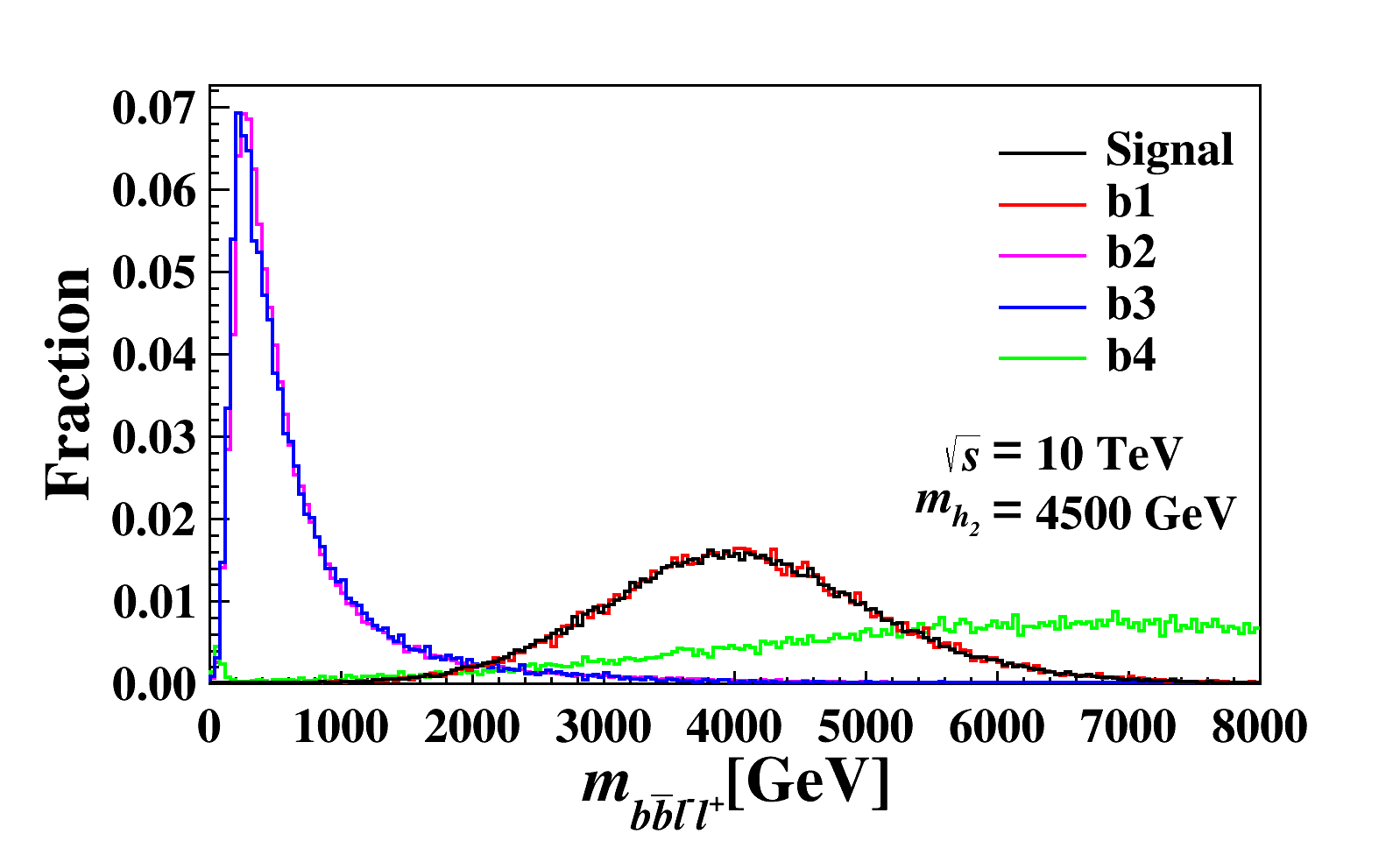}
\hspace{-6mm}
\includegraphics[width=0.33\linewidth]{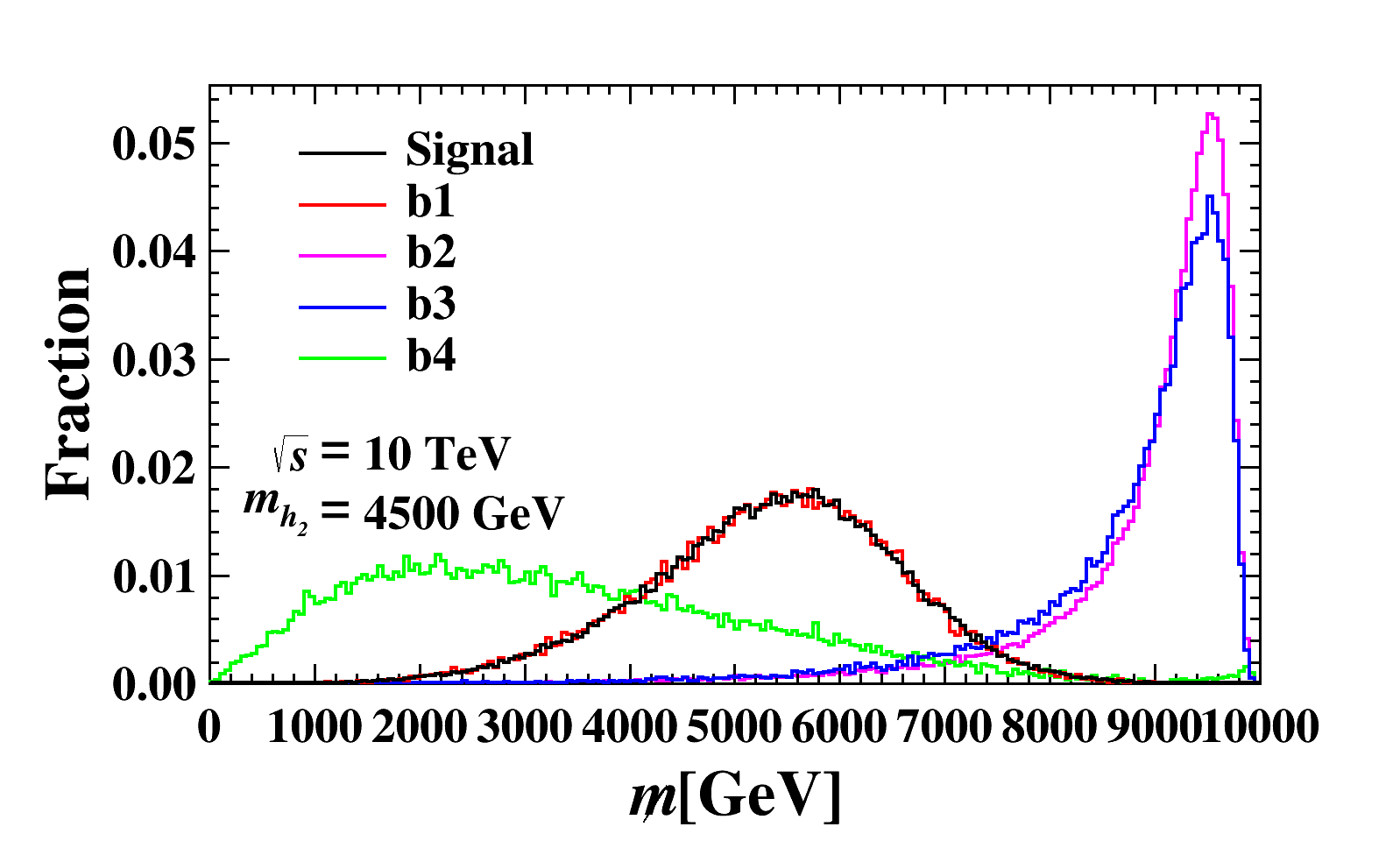}
\caption{Normalized distributions of the observables $m_{b\bar b}$ (left), $m_{b\bar b\ell^-\ell^+}$ (middle), and $m\mkern-10.5mu/$ (right), for all the other six benchmark points from top to bottom: $m_{h_2}=1000~\mathrm{GeV}$ with $\sqrt{s}=3$ TeV, and $m_{h_2}=700,1000,1500,2000,4500~\mathrm{GeV}$ with $\sqrt{s}=10$ TeV.}
\label{fig:app-reco-10tev-0p7}
\end{figure}

We show the normalized distributions of the observables
$m_{b\bar{b}}$, $m_{b\bar{b}\ell^-\ell^+}$, and $\slashed{m}$ in Fig. \ref{fig:app-reco-10tev-0p7} for the other six benchmark
points as mentioned above.

\section{Cross sections after each step}
\label{app:cuts}

We show the cross sections for both signal and background processes after each step: initial production (denoted as ``ini.''), pre-selection cuts (denoted as ``pre-sel.''), and selection cuts (denoted as ``sel.''). Results for the two benchmark points with $\sqrt{s}=3~\mathrm{TeV}$ are shown in Table \ref{tab:sel3TeV}, and results for the six benchmark points with $\sqrt{s}=10~\mathrm{TeV}$ are shown in Table \ref{tab:sel10TeV}

\begin{table}[h]
\centering
\renewcommand{\arraystretch}{1.3}
\begin{tabular}{cccccccc}
\hline
\hline
$m_{h_2}$ [GeV] & cuts  & $\sigma_{\mathrm{sig}}$~[fb] & $\sigma_{\mathrm{bkg}}$~[fb] & $\sigma_{\mathrm{b1}}$~[fb] & $\sigma_{\mathrm{b2}}$~[fb] & $\sigma_{\mathrm{b3}}$~[fb] & $\sigma_{\mathrm{b4}}$~[fb] \\
\hline
\multirow{3}{*}{700} & ini. & 6.75$\kappa$ & 8.16$\xi$+5.5 & 8.16$\xi$ & 0.39 & 2.71 & 2.40   \\
  & pre-sel. & 4.13$\kappa$ & 1.12$\xi$+0.96 & 1.12$\xi$ & 0.18 & 0.34 & 0.44 \\
  & sel. & 1.58$\kappa$ & 0.014$\xi$+0.0051 & 0.014$\xi$ & 0.0045 & 0.00035 & 0.00025 \\
\hline
\multirow{3}{*}{1000} & ini. & 3.90$\kappa$ & 4.68$\kappa$+5.5 & 4.68$\kappa$ & 0.39 & 2.71 & 2.40 \\
  & pre-sel. & 2.49$\kappa$ & 0.72$\xi$+0.96 & 0.72$\xi$ & 0.18 &  0.34 & 0.44\\
  & sel. & 0.86$\kappa$ & 0.007$\xi$+0.0028 & 0.007$\xi$ & 0.0022 & 0.00016 & 0.00044   \\
\hline
\hline
\end{tabular}
\caption{Cross sections for signal and background processes after each step (initial production, pre-selection cuts, and selection cuts) for benchmark points with $\sqrt{s} = 3$ TeV.}
\label{tab:sel3TeV}
\end{table}

\begin{table}[h]
\centering
\renewcommand{\arraystretch}{1.3}
\begin{tabular}{cccccccc}
\hline
\hline
$m_{h_2}$ [GeV] & cuts  & $\sigma_{\mathrm{sig}}$~[fb] & $\sigma_{\mathrm{bkg}}$~[fb] & $\sigma_{\mathrm{b1}}$~[fb] & $\sigma_{\mathrm{b2}}$~[fb] & $\sigma_{\mathrm{b3}}$~[fb] & $\sigma_{\mathrm{b4}}$~[fb] \\
\hline
\multirow{3}{*}{700} & ini. & 20.12$\kappa$ & 24.18$\xi$+11.4 & 24.18$\xi$ & 1.39 & 9.6 & 0.41  \\
  & pre-sel. & 10.58$\kappa$ & 3.24$\xi$+1.37 & 3.24$\xi$ & 0.44 & 0.90 & 0.03 \\
  & sel. & 3.74$\kappa$ & 0.035$\xi$+0.00987 & 0.035$\xi$ & 0.0092 & 0.00067 & $-$  \\
\hline
\multirow{3}{*}{1000} & ini. & 16.15$\kappa$ & 19.42$\xi$+11.4 & 19.42$\xi$ & 1.39 & 9.60 & 0.41 \\
  & pre-sel. & 9.30$\kappa$ & 2.83$\xi$+1.37 & 2.83$\xi$ & 0.44 & 0.90 & 0.03  \\
  & sel. & 3.02$\kappa$ & 0.038$\xi$+0.00625 & 0.038$\xi$ & 0.0061 & 0.00015 & $-$  \\
\hline
\multirow{3}{*}{1500} & ini. & 11.76$\kappa$ & 14.13$\xi$+11.4 & 14.13$\xi$ & 1.39 & 9.6 & 0.41\\
  & pre-sel. & 7.27$\kappa$ & 2.19$\xi$+1.37 & 2.19$\xi$ & 0.44 & 0.90 & 0.03 \\
  & sel. & 2.96$\kappa$ & 0.032$\xi$+0.00850 & 0.032$\xi$ & 0.0084 & 0.00010 & $-$ \\
\hline
\multirow{3}{*}{2000} & ini. & 8.70$\kappa$ & 10.46$\xi$+11.4 & 10.46$\xi$ & 1.39 & 9.6 & 0.41 \\
  & pre-sel. & 5.40$\kappa$ & 1.62$\xi$+1.37 & 1.62$\xi$ & 0.44 & 0.90 & 0.03  \\
  & sel. & 2.11$\kappa$ & 0.025$\xi$+0.00752 & 0.025$\xi$ & 0.0074 & 0.00012 & $-$ \\
\hline
\multirow{3}{*}{3000} & ini. & 4.93$\kappa$ & 5.93$\xi$+11.4 & 5.93$\xi$ & 1.39 & 9.6 & 0.41  \\
  & pre-sel. & 2.80$\kappa$ & 0.85$\xi$+1.37 & 0.85$\xi$ & 0.44 & 0.90 & 0.03  \\
  & sel. & 1.13$\kappa$ & 0.017$\xi$+0.00707 & 0.017$\xi$ & 0.0069 & 0.00017 & $-$  \\
\hline
\multirow{3}{*}{4500} & ini. & 1.98$\kappa$ & 2.38$\xi$+11.4 & 2.38$\xi$ & 1.39 & 9.6 & 0.41  \\
  & pre-sel. & 0.67$\kappa$ & 0.21$\xi$+1.37 & 0.21$\xi$ & 0.44 & 0.90 & 0.03  \\
  & sel. & 0.28$\kappa$ & 0.007$\xi$+0.00493 & 0.007$\xi$ & 0.0038 & 0.00083 & 0.00030  \\
\hline
\hline
\end{tabular}
\caption{Cross sections for signal and background processes after each step (initial production, pre-selection cuts, and seection cuts) for benchmark points with $\sqrt{s} = 10$ TeV. ``$-$'' means this background process is negligible with $L=10~\mathrm{ab}^{-1}$.}
\label{tab:sel10TeV}
\end{table}

\newpage
\acknowledgments
We thank Zilong Ding, Jin Sun, Yiheng Xiong, and Rui Zhang for helpful discussions. Q.L. and Y.N.M. are supported by the National Natural Science Foundation of China under Grant No. 12205227. K.W. is supported by the National Natural Science Foundation of China under Grant No. 11905162, the Excellent Young Talents Program of the Wuhan University of Technology under Grants No. 40122102 and No. 40122103, and the research program of the Wuhan University of Technology under Grants No. 2020IB024 and No. 104972025KFYjc0101. The simulation and analysis work of this article was completed with the computational cluster provided by the Theoretical Physics Group at the Department of Physics, School of Physics and Mechanics, Wuhan University of Technology.

\paragraph{Note added.}
Following standard practice in high energy physics, authors are listed in strict alphabetical order by surname.
This ordering should not be interpreted as indicating any ranking of contribution, seniority, or leadership.
All authors contributed equally to this work.


\bibliographystyle{JHEP}
\bibliography{paperfinal}

\end{document}